\documentclass[12pt,preprint]{aastex}

\usepackage{graphicx}

\begin{document}

% Title, etc. {{{

\shorttitle{Damping in KAW Turbulence} 

\shortauthors{Smith \& Terry}

\title{Damping of Electron Density Structures and Implications for Interstellar
Scintillation}
  
\author{K.W. Smith and P.W. Terry}

\affil{Center for Magnetic Self Organization in Laboratory and Astrophysical
Plasmas and Department of Physics, University of Wisconsin-Madison, Madison, WI
53706} 

\email{kwsmith1@wisc.edu}
%}}}

\begin{abstract}%{{{

The forms of electron density structures in kinetic Alfv\'en wave turbulence
are studied in connection with scintillation. The focus is on small scales $L
\sim 10^8-10^{10}$ cm where the Kinetic Alfv\'en wave (KAW) regime is active in
the interstellar medium, principally within turbulent HII regions around bright
stars. MHD turbulence converts to a KAW cascade, starting at 10 times the ion
gyroradius and continuing to smaller scales.  These scales are inferred to
dominate scintillation in the theory of Boldyrev \emph{et al.}
\citep{boldyrev03a,boldyrev03b,boldyrev05,boldyrev06}.  From numerical
solutions of a decaying kinetic Alfv\'en wave turbulence model, structure
morphology reveals two types of localized structures, filaments and sheets, and
shows that they arise in different regimes of resistive and diffusive damping. 
Minimal resistive damping yields localized current filaments that form out of
Gaussian-distributed initial conditions.  When resistive damping is large
relative to diffusive damping, sheet-like structures form.  In the filamentary
regime, each filament is associated with a non-localized magnetic and density
structure, circularly symmetric in cross section.  Density and magnetic fields
have Gaussian statistics (as inferred from Gaussian-valued kurtosis) while
density gradients are strongly non-Gaussian, more so than current.  This
enhancement of non-Gaussian statistics in a derivative field is expected since
gradient operations enhance small-scale fluctuations.  The enhancement of
density gradient kurtosis over current kurtosis is not obvious, yet it suggests
that modest fluctuation levels in electron density may yield large
scintillation events during pulsar signal propagation in the interstellar
medium.  In the sheet regime the same statistical observations hold, despite
the absence of localized filamentary structures.  Probability density functions
are constructed from statistical ensembles in both regimes, showing clear
formation of long, highly non-Gaussian tails.

\end{abstract}

%}}}

\section{Introduction}%{{{

Models of scintillation have a long history. Many
\citep{lee75a,lee75b,sutton71} carry an implicit or explicit assumption of
Gaussian statistics, applying to either the electron density field itself or
its autocorrelation function (herein referred to as ``Gaussian Models'').
\citet{lee75a} is a representative approach. The statistics of the two-point
correlation function of the index of refraction $\epsilon(\mathbf{r})$,
$A(\rho) = \int{dz' \left< \epsilon(x,z) \epsilon(x+\rho,z') \right>}$
determines, among other effects, the scaling of pulsar signal width $\tau$ with
dispersion measure $DM$.  The index of refraction $\epsilon(\mathbf{r})$ is a
function of electron density fluctuation $n(\mathbf{r})$. The quantity $A(0)$
enters the equations, representing the second moment of the index of
refraction.  If the distribution function of $\epsilon(\mathbf{r})$ has no
second-order moment (as in a L\'evy distribution) $A(0)$ is undefined.  The
assumption of Gaussian statistics leads to a scaling of $\tau \sim DM^2$, which
contradicts observation for pulsars with $DM > 30$ cm$^{-3}$ pc.  For these
distant pulsars, $\tau \sim DM^4$ \citep{sutton71,boldyrev03a,boldyrev03b}. 

To explain the anomalous $DM^4$ scaling, \citet{sutton71} argued that the
pulsar signal encounters strongly scattering turbulent regions for longer lines
of sight, essentially arguing that the statistics, as sampled by a pulsar
signal, are nonstationary.  Considering the pulse shape in time,
\citet{williamson72,williamson73,williamson74} is unable to match observations
with a Gaussian Model of scintillation unless the scattering region is confined
to $1/4$ of the line of sight between the pulsar and Earth.  These assumptions
may have physical basis, since the ISM may not be statistically stationary,
being composed of different regions with varying turbulence intensity
\citep{boldyrev05}.  

The theory of \citet{boldyrev03a,boldyrev03b,boldyrev05,boldyrev06} takes a
different approach to explain the anomalous $DM^4$ scaling by considering
L\'evy statistics for the density difference (defined below).  L\'evy
distributions are characterized by long tails with no defined moments greater
than first-order [i.e., $A(0)$ is undefined for a L\'evy distribution].  The
theory recovers the $\tau \sim DM^4$ relation with a statistically stationary
electron density field.  This theory also does not constrain the scattering
region to a fraction of the line-of-sight distance.

The determinant quantity in the theory of Boldyrev \emph{et al.} is the density
difference, $\Delta n = n(\mathbf{x}_1, z) - n(\mathbf{x}_2, z)$. According to
this model, if the distribution function of $\Delta n$ has a power-law decay as
$|\Delta n| \rightarrow \infty$ and has no second moment, then it is possible
to recover the $\tau \sim DM^4$ scaling \citep{boldyrev03b}.  Assuming
sufficiently smooth fluctuations, $\Delta n$ can be expressed in terms of the
density gradient, $\mathbf{\sigma}(z)$: $n(\mathbf{x}_1) - n(\mathbf{x}_2)
\simeq \mathbf{\sigma}(z) \cdot (\mathbf{x}_1 - \mathbf{x}_2)$.  Perhaps more
directly, the density gradient enters the ray tracing equations [Eqns. (7) in
\citet{boldyrev03a}], and is seen to be central to determining the resultant
pulsar signal shape and width.  This formulation of a scintillation theory does
not require the distribution of $\Delta n$ to be Gaussian or to have a
second-order moment.

The notion that the density difference is characterized by a L\'evy
distribution is a constraint on dynamical models for electron density
fluctuations in the ISM.  Consequently the question of how a L\'evy
distribution can arise in electron density fluctuations assumes considerable
importance in understanding the ISM.

Previous work has laid the groundwork for answering this question.  It has been
established that electron density fluctuations associated with interstellar
magnetic turbulence undergo a significant change in character near the scale
$10 \rho_s$, where $\rho_s$ is the ion sound gyroradius \citep{ter01}.  At
larger scales, electron density is passively advected by the turbulent flow of
an MHD cascade mediated by nonlinear shear Alfv\'en waves \citep{goldreich95}. 
At smaller scales the electron density becomes compressive and the turbulent
energy is carried into a cascade mediated by kinetic Alfv\'en waves (KAW)
\citep{ter01}.  The KAW cascade brings electron density into equipartition with
the magnetic field, allowing for a significant increase in amplitude.  The
conversion to a KAW cascade has been observed in numerical solutions of the
gyrokinetic equations \citep{howes06}, and is consistent with observations from
solar wind turbulence \citep{harmon05,bale05,leamon98}.  Importantly, it puts
large amplitude electron density fluctuations (and large amplitude density
gradients) at the gyroradius scale ($\sim 10^8-10^{10}$ cm), a desirable set of
conditions for pulsar scintillation \citep{boldyrev06}.  It is therefore
appropriate to consider whether large-amplitude non-Gaussian structures can
arise in KAW turbulence. 

This question has been partially answered in a study of current filament
formation in decaying KAW turbulence \citep{terry-smith07,terry-smith08}. 
In numerical solutions to a two-field model with broadband Gaussian initial
conditions large amplitude current filaments spontaneously arose.  Each
filament was associated with a large-amplitude electron density structure,
circular in cross-section, that persisted in time.  These electron density
structures were not as localized as the corresponding current filaments, but
were coherent and not mixed by surrounding turbulence.  The observation of
large amplitude current filaments is similar to the large-amplitude vortex
filaments found in decaying 2D hydrodynamic turbulence \citep{mcw84}. 
Counterparts of such structures in 3D are predicted to be the dominant
component for higher order structure functions \citep{she94}. 

\citet{terry-smith07} proposed that a nonlinear refractive magnetic shear
mechanism prevents the structures from mixing with turbulence.  Radial shear in
the azimuthally directed magnetic field associated with each large-amplitude
current filament decreases the radial correlation length of the turbulent
eddies and enhances the decorrelation rate.  Eddies are unable to persist long
enough to penetrate the shear boundary layer and disrupt the structure core. 
The structure persistence mechanism allows large-amplitude fluctuations to
persist for many eddy-turnover times.  As the turbulent decays these structures
eventually dominate the statistics of the system.  The spatial structure of the
density field associated with localized circularly symmetric current filaments
was shown from analytical theory \citep{terry-smith07} to yield a
L\'evy-distributed density gradient field.  The kurtosis for the current field
was significantly larger than the Gaussian-valued kurtosis of 3, indicating
enhanced tails.  The electron density and magnetic field kurtosis values were
not significantly greater than 3.  However, just as the current is non-Gaussian
when the magnetic field is not, it is expected that numerical solutions should
show non-Gaussian behavior for the density gradient.  In the present paper,
density gradient statistics are measured and found to be non-Gaussian.  Rather
than relying on kurtosis values alone, the probability density functions (PDFs)
are computed from ensembles of numerical solutions, showing non-Gaussian PDFs
for the density gradient field.

The previous studies of filament generation in KAW turbulence leave significant
unanswered questions relating to structure morphology and its effect on
scintillation.  It is well established that MHD turbulence admits structures
that are both filament-like and sheet-like.  Can sheet-like structures arise in
KAW turbulence?  If so, what are the conditions or parameters favoring one type
of structure versus the other?  If sheet-like structures dominate in some
circumstances, what are the statistics of the density gradient?  Can they be
sufficiently non-Gaussian to be compatible with pulsar scintillation scaling?
It is desirable to consider such questions prior to calculation of rf wave
scattering properties in the density gradient fields obtained from numerical
solutions.

In this paper, we show that both current filaments and current sheet structures
naturally arise in numerical solutions of a decaying KAW turbulence model. 
Each has a structure of the same type and at the same location in the electron
density gradient.  These structures become prevalent as the numerical solutions
progress in time, and each is associated with highly non-Gaussian PDFs.
Moreover, we show that small-scale current filaments and current sheets, along
with their associated density structures, are highly sensitive to the magnitude
of resistive damping and diffusive damping of density fluctuations.  Current
filaments persist provided that resistivity $\eta$ is small; similarly,
electron density fluctuations and gradients are diminished by large diffusive
damping in the electron continuity equation.  The latter results from
collisions assuming density fluctuations are subject to a Fick's law for
diffusion.  The magnitude of the resistivity affects (1) whether current
filaments can become large in amplitude, (2) their spatial scale, and (3) the
preponderance of these filaments as compared to sheets.  The magnitude of the
diffusive damping parameter, $\mu$, similarly influences the amplitude of
density gradients and, to a lesser degree, influences the extent to which
electron density structures are non-localized.  

In the ISM resistive and diffusive damping become important near resistive
scales.  However, it is well known that collisionless damping effects are also
present \citep{lysak96,bale05}, and quite possibly dominate over collisional
damping in larger scales near the ion Larmor radius.  The collisional damping
in the present work is understood as a heuristic approach that facilitates
analysis of the effects of different damping regimes on the statistics of
electron density fluctuations.  By varying the ratio of resistive and diffusive
damping we can, as suggested above, control the type of structure present in
the turbulence.  This allows us to isolate and study the statistics associated
with each type of structure.  It also allows us to assess and examine the type
of environment conducive to formation of the structure.  We consider regimes
with large and small damping parameters, enabling us to explore damping effects
on structure formation across a range from inertial to dissipative.  Future
work will address collisionless damping in greater detail.

\subsection{Background Considerations for Structure Formation}

The coherent structures observed in numerical solutions of decaying KAW
turbulence, whether elongated sheets or localized filaments, are similar to
structures observed in decaying MHD turbulence, as in \citet{kinney95}.  In
that work, the flow field initially gives rise to sheet-like structures. After
selective decay of the velocity field energy, the system evolves into a state
with sheets and filaments.  During the merger of like-signed filaments,
large-amplitude sheets arise, limited to the region between the merging
filaments.  These short-lived sheets exist in addition to the long-lived sheets
not associated with the merger of filaments.  In the two-field KAW system,
however, there is no flow; the sheet and filament generation is due to a
different mechanism, of which the filament generation has previously been
discussed \citep{terry-smith07}.

Other work \citep{biskamp89,politano89} observed the spontaneous generation of
current sheets and filaments in numerical solutions, with both Orszag-Tang
vortex and randomized initial conditions.  These 2D reduced MHD numerical
solutions modeled the evolution of magnetic flux and vorticity with collisional
dissipation coefficients $\eta$, the resistivity, and $\nu$, the kinematic
viscosity.  The magnetic Prandtl number, $\nu / \eta$, was set to unity.  These
systems are incompressible and not suitable for modeling the KAW system we
consider here -- they do however illustrate the ubiquity of current sheets and
filaments, and serve as points of comparison.  For Orszag-Tang-like initial
conditions with large-scale flux tubes smooth in profile, current sheets are
preferred at the interfaces between tubes.  Tearing instabilities can give rise
to filamentary current structures that persist for long times, but the
large-scale and smoothness of flux tubes do not give rise to strong current
filaments localized at the center of the tubes.  To see this, consider a given
flux tube, and model it as cylindrically symmetric and monotonically decreasing
in $r$ with characteristic radial extent $a$,

\begin{equation}
\psi(r) = \psi_0 \left( 1 - (\frac{r}{a})^2 \right),
\end{equation}

\noindent{ for $0 \leq r \leq a$. The current is localized at the center with
magnitude}

\begin{equation}
J = - 4 \frac{\psi_0}{a^2}.
\end{equation}

\noindent{ Thus flux tubes with large radial extent $a$ have a corresponding
small current filament at their center. Hence, initial conditions dominated by
a few large-scale flux tubes are not expected to have large amplitude current
filaments at the flux tube centers, but favor current sheet formation and
filaments associated with tearing instabilities in those sheet regions.  At X
points current sheet folding and filamentary structures can arise \citep[see,
e.g.][Fig.\ 10]{biskamp89}, but these regions are small in area compared to the
quiescent flux-tube regions.  Note that if, instead of Orszag-Tang-like initial
conditions, the initial state is random, one expects some regions with flux
tubes that have $a$ small, and therefore a sizable current filament at the
center.}

Consider now the effect of comparatively large or small $\eta$.  In the case of
large $\eta$, the central region of a flux tube is smoothed by the collisional
damping, thus having a strong suppressive effect on the amplitude of the
current filament associated with such a flux tube.  Large-amplitude current
structures are localized to the interfaces between flux tubes.  In the process
of mergers between like-signed filaments (and repulsion between unlike-signed),
large current sheets are generated at these interfaces, similar to the
large-amplitude sheets generated in MHD turbulence during mergers
\citep{kinney95}.  For small $\eta$, relatively little suppression of isolated
current filaments should occur; if these filaments are spatially separated
owning to the buffer provided by their associated flux tube, they can be
expected to survive a long time and only be disrupted upon the merger with
another large-scale flux tube.  Large $\eta$, then, allows current sheets to
form at the boundaries between flux tubes while suppressing the
spatially-separated current filaments at flux tube centers. Small $\eta$ allows
interface sheets and spatially separated filaments to exist.

These simple arguments suggest that the evolution of the large-amplitude
structures and their interaction with turbulence is thus strongly influenced by
the damping parameters.  As such, the magnitudes of the damping parameters are
expected to affect the resultant pulsar scintillation scalings.  The present
paper considers the effect of variations of these damping parameters, $\eta$
and $\mu$. In the KAW model, the (unnormalized) resistivity takes the form
$\eta = m_e \nu_e / n e^2$ and the density diffusion coefficient is $\mu =
\rho_e^2 \nu_e$, where $m_e$ is the electron mass, $\nu_e$ is the electron
collision frequency, $n$ is the electron density, $e$ is the electron charge,
and $\rho_e = v_{Te} / \omega_{ce}$, with $v_{Te}$ the electron thermal
velocity, and $\omega_{ce}$ the electron gyrofrequency.  The ratio of these
terms, $c^2 \eta / 4\pi \mu = 2 / \beta$, where $\beta = 8 \pi n k T / B^2$ is
the ratio of plasma to magnetic pressures.  When we vary this ratio, as we will
do in the numerical solutions presented here, we have in mind that we are
representing regions of different $\beta$.  However, as a practical matter in
the numerical solutions, we must vary the damping parameters independently of
the variation of $\beta$, since the kinetic Alfv\'en wave dynamics require a
small $\beta$ to propagate.  For the warm ionized medium, typical parameters
are $T_e = 8000\;\mathrm{K}$, $n = 0.08\;\mathrm{cm}^{-3}$, $|B| =
1.4\;\mu\mathrm{G}$, $\delta B = 5.0\;\mu\mathrm{G}$ \citep{ferriere01}. With
these parameters, the plasma $\beta$ formally ranges from $0.05-1.2$, spanning
a range of plasma magnetization.

We present the results of numerical solutions of decaying KAW turbulence to
ascertain the effect of different damping regimes on the statistics of the
fields of interest, in particular the electron density and electron density
gradient.  In the $\eta \ll \mu$ regime (using normalized parameters), previous
work \citep{crad91} had large-amplitude current filaments that were strongly
localized with no discernible electron density structures ($\mu$ was large to
preserve numerical stability).  This regime is unable to preserve density
structures or density gradients.  The numerical solutions presented here have
$\eta \sim \mu$ and $\eta \gg \mu$; in each limit the damping parameters are
minimized so as to allow structure formation to occur, and are large enough to
ensure numerical stability for the duration of each numerical solution.  We
investigate the statistics of both filaments and sheets in the context of
scintillation in the warm ionized medium.

The paper is arranged as follows:  section \ref{sec:KAW-model} gives an
overview of the KAW model and normalizations, its regime of validity, and its
dispersion relation. Section \ref{sec:numerical-solution} discusses the
numerical method used and the field initializations.  The negligible effect of
initial cross-correlation between fields is discussed.  Results for the two
damping regimes are given in section \ref{sec:results}, where the type of
structures that form, whether sheets-and-filaments or predominantly sheets, are
seen to be dependent on the values of $\eta$ and $\mu$.  PDFs from ensemble
numerical solutions are presented in section \ref{sec:PDFs}, illustrating the
strongly non-Gaussian statistics in the electron density gradient field for
both the $\eta \sim \mu$ and $\eta \gg \mu$ regimes. This suggests that
non-Gaussian electron-density gradients are robust to variation in $\eta$, as
long as the overall damping in the continuity equation is not too large.  Some
discussions regarding the limitations of numerical approximation for this work
and possible enhancements--particularly a model that addresses driven KAW
turbulence--are given in concluding remarks.

%}}}

\section{Kinetic Alfv\'{e}n Wave Model}%{{{
\label{sec:KAW-model}

The kinetic Alfv\'en wave (KAW) model used in this paper is the same model used
in theories of pulsar scintillation through the ISM \citep{terry-smith07,
terry-smith08} and in earlier work \citep{crad91}.  It is a reduced, two-field,
small-scale limit of a more general reduced three-field MHD system
\citep{haz83,rahman83,fernandez97} that accounts for electron dynamics parallel
to the magnetic field.

The 3-field model applies to large and small-scale fluctuations as compared to
$\rho_s$, the ion gyroradius evaluated at the electron temperature.  In
large-scale strong turbulence magnetic and kinetic fluctuations are in
equipartition, with electron density passively advected.  In the limit of small
spatial scales ($\leq 10 \rho_s$) the roles of kinetic and internal
fluctuations are reversed -- magnetic fluctuations are in equipartition with
density fluctuations, and kinetic energy experiences a go-it-alone cascade
without participating in the magnetic-internal energy interaction.  The
shear-Alfv\'en physics at large scale is supplanted by kinetic-Alfv\'en physics
at small scale \citep{ter01}.

In the Boldyrev \emph{et al.} theory, the length scales that dominate
scintillation for pulsars with $DM > 30 \;\mathrm{pc}\;\mathrm{cm}^{-3}$ are
small, around $10^8-10^{10}\;\mathrm{cm}$.  This motivates our focus on the
small-scale regime of the more general 3-field system.  The dominant
interactions are between magnetic and internal fluctuations, via kinetic
Alfv\'en waves.  In these waves, electron density gradients along the magnetic
field act on an inductive electric field in Ohm's law.  The electron continuity
equation serves to close the system.  The normalized equations are

\begin{equation}
\partial_t \psi = \nabla_{\|} n + \eta_0 J - \eta_2 \nabla^2 J,
\label{psi-eqn}
\end{equation}
\begin{equation}
\partial_t n    = - \nabla_{\|} J + \mu_0 \nabla^2 n - \mu_2 \nabla^2 \nabla^2 n,
\label{den-eqn}
\end{equation}
\begin{equation}
\nabla_{\|} = \partial_z + \nabla\psi \times \mathrm{z} \cdot \nabla,
\end{equation}
\begin{equation}
J = \nabla^2 \psi,
\end{equation}

\noindent{ with $\psi = (C_s/c)e A_z/T_e$, the normalized parallel component of
the vector potential and $n = (C_s/V_A)\tilde{n}/n_0$ the normalized electron
density.  The normalized resistivity is $\eta_0 = (c^2/4\pi V_A \rho_s)
\eta_{sp}$, with $\eta_{sp}$ the Spitzer resistivity, given in the
introduction.  The normalized diffusivity is $\mu = \rho_e^2 \nu_e /\rho_s
V_A$.  The time and space normalizations are $\tau_A = \rho_s / V_A$ and
$\rho_s = C_s / \Omega_i$. Here $C_s = (T_e/m_i)^{1/2}$ is the ion acoustic
velocity, $V_A = B / (4\pi m_i n_0)^{1/2}$ is the Alfv\'en speed, and $\Omega_i
= eB/m_i c$ is the ion gyrofrequency.  Electron density diffusion is presumed
to follow Fick's law; more detailed damping would necessarily consider kinetic
effects and cyclotron resonances.  The $\eta_2$ and $\mu_2$ terms
(hyper-resistivity and hyper-diffusivity) are introduced to mitigate
large-scale Fourier-mode damping by the linear diffusive terms. Throughout the
remainder of the paper, we drop the $2$ subscript from $\eta_2$ and $\mu_2$ and
refer to the hyper-dissipative terms as $\eta$ and $\mu$.}  

Three ideal invariants exist: total energy $E = \int d^2x [(\nabla \psi)^2 +
\alpha n^2]$; flux $F = \int d^2x\; \psi^2$ and cross-correlation $H_c = \int
d^2 x\; n\psi$.  Energy cascades to small scale (large $k$) while the flux and
cross-correlation undergo an inverse cascade to large-scale (small $k$)
\citep{fernandez97}. The inverse cascades require the initialized spectrum to
peak at $k_0 \neq 0$ to allow for buildup of magnetic flux at large-scales for
later times.

Linearizing the system yields a (dimensional) dispersion relation $\omega
= V_A k_z k_\perp \rho_s$.  The mode combines perpendicular oscillation
associated with a finite gyroradius with fluctuations along a mean field
($z$-direction).  The oscillating quantities are magnetic field and density,
out of phase by $\pi/2$ radians.

In the limit of strong mean field, quantities along the mean field
($z$-direction) equilibrate quickly, which allows $\partial / \partial z
\rightarrow 0$, or $k_z \rightarrow 0$.  Kinetic Alfv\'en waves still
propagate, as long as there are a broad range of scales that are excited, as in
fully developed turbulence.  As $k_z \rightarrow 0$, all gradients are
localized to the plane perpendicular to the mean field.  Presuming a
large-scale fluctuation at characteristic wavenumber $\mathbf{k}_0$,
smaller-scale fluctuations propagate linearly along this larger-scale
fluctuation so long as their characteristic scale $k$ satisfies $k \gg k_0$.
In this reduced, two dimensional system, the above dispersion relation is
modified to be $\omega = V_A (\mathbf{b}_{k_0} \cdot \mathbf{k} / B) k \rho_s$
which is still Alfv\'enic but with respect to a perturbed large-scale amplitude
perpendicular to the mean field.  Relaxing the scale separation criterion
yields $\omega \propto k^2$ for the general case.

%}}}

\section{Numerical Solution Method}%{{{
\label{sec:numerical-solution}

We evolve Eqs.\ (\ref{psi-eqn}) and (\ref{den-eqn}) in a 2D periodic box, size
$[2\pi] \times [2\pi]$ on a mesh of resolution $512 \times 512$.  The $\psi$
and $n$ scalar fields are evolved in the Fourier domain, with the
nonlinearities advanced pseudospectrally and with full $2/3$ dealiasing in each
dimension \citep{orszag71}.  The diffusive and resistive terms normally
introduce stiffness into the equations; using an integrating factor removes any
stability constraints stemming from these terms.  Following the scheme outlined
in \citet{canuto90}, we start with the semi-discrete formulation of Eqs.\ 
(\ref{psi-eqn}) and (\ref{den-eqn}):

\begin{equation}
\frac{d \psi_k}{d t} = -\eta k^4 \psi_k + \mathcal{F} \left[ \nabla_{\|} n \right]
\label{psi_discrete}
\end{equation}

\begin{equation}
\frac{d n_k}{d t} = -\mu k^4 n_k - \mathcal{F} \left[ \nabla_{\|} J \right],
\label{den_discrete}
\end{equation}

\noindent{ where $\mathcal{F}[\cdot]$ denotes the discrete Fourier transform.
We do not explicitly expand the nonlinear terms as they will be integrated
separately.  The hyper-damping terms (proportional to $k^4$) are included
above.  Damping terms corresponding to the Laplacian operator (proportional to
$k^2$) are not included in this section for clarity, but are trivial to
incorporate.  Equations \ref{psi_discrete} and \ref{den_discrete} can be put in
the form }

\begin{equation}
\frac{d}{dt} \left[e^{\eta k^4 t} \psi_k \right] = e^{\eta k^4 t} \mathcal{F} \left[ \nabla_{\|} n \right]
\end{equation}

\begin{equation}
\frac{d}{dt} \left[e^{\mu k^4 t} n_k \right] = - e^{\mu k^4 t} \mathcal{F}
\left[ \nabla_{\|} J \right].
\end{equation}

\noindent{
A second-order Runge-Kutta scheme for the $\psi_k$ difference equation is
}

\begin{equation}
\psi_k^{m+1/2} = e^{- \eta k^4 \Delta t/2} \left[ 
                            \psi_k^m + \Delta t/2
                            \mathcal{F}\left[ \nabla_{\|} n^m \right] 
                            \right]
\end{equation}

\begin{equation}
\psi_k^{m+1} = e^{- \eta k^4 \Delta t} \left[ 
                            \psi_k^{m+1/2} + \Delta t
                            \mathcal{F}\left[ \nabla_{\|} n^{m+1/2} \right] 
                            \right]
\end{equation}

\noindent{ with a similar form for the $n_k$ scheme.}

\subsection{Initial conditions}

The $\psi_k$ and $n_k$ fields are initialized such that the energy spectra are
broad-band with a peak near $k_0 \sim 6-10$ and a power law spectrum for $k >
k_0$.  The falloff in $k$ is predicted to be $k^{-2}$ for small-scale
turbulence. \citet{crad91} use $k^{-3}$, between the current-sheet limit of
$k^{-4}$ and the kinetic-Alfv\'en wave strong-turbulence limit of $k^{-2}$.
The numerical solutions considered here have either $k^{-2}$ or $k^{-3}$.  The
only qualitative difference between the two spectra is the scale at which
structures initially form.  The $k^{-2}$ spectra has more energy at smaller
scales, leading to smaller characteristic structure size.  After a few tens of
Alfv\'en times these smaller-scale structures merge and the system resembles
the initial $k^{-3}$ spectra.

The $n_k$ and $\psi_k$ phases can be either cross-correlated or uncorrelated.
By cross-correlated we mean that the phase angle for each Fourier component of
the $n_k$ and $\psi_k$ fields are equal.  In general,

\begin{equation}
n_k = |A_k| e^{i\theta_1}, \qquad \psi_k = |B_k| e^{i \theta_2},
\end{equation}

\noindent{ where $|A_k|$ and $|B_k|$ are the Fourier component's amplitude, set
according to the spectrum power-law.  For cross-correlated initial conditions,
$\theta_1 = \theta_2$ for all $\mathbf{k}$ at the initial time.  For
uncorrelated initialization, there is no phase relation between corresponding
Fourier components of the $n_k$ and $\psi_k$ fields.}

\citet{crad91} focused on the formation and longevity of current filaments in a
turbulent KAW system.  To preserve small-scale structure in the current
filaments, these numerical solutions set $\eta = 0$ and had $\mu \sim 10^{-3}$,
with a resolution of $128 \times 128$, corresponding to a $k_{max}$ of 44.
Large-amplitude density structures that would have arisen were damped to
preserve numerical stability up to an advective instability time of a few
hundred Alfv\'en times, for the parameter values therein.  The numerical
solutions presented here explore a range of parameter values for $\eta$ and
$\mu$.  They make use of hyper-diffusivity and hyper-resistivity of appropriate
strengths to preserve structures in $n$, $B$ and $J$.  An advective instability
is excited after $\sim 10^2$ Alfv\'en times if resistive damping is negligible.
The $\eta = 0$ solutions--not presented here due to their poor resolution of
small-scale structures--see large-amplitude current filaments arise, but they
can be poorly resolved at this grid spacing. With no resistivity, the finite
number of Fourier modes cannot resolve arbitrarily small structures without
Gibbs phenomena resulting and distorting the current field.  

We have found through experience that small hyper-resistivity and small
hyper-diffusivity preserve large-amplitude density structures and their spatial
correlation with the magnetic and current structures, while preventing the
distortion resulting from poorly-resolved current sheets and filaments. They
allow the numerical solutions to run for arbitrarily long times, and the
effects of structure mergers become apparent. These occur on a longer timescale
than the slowest eddy turnover times.  The results presented here will consider
two regimes of parameter values, the $\eta \approx \mu$ and $\eta \gg \mu$
regimes.  The effect of cross-correlated and uncorrelated initial conditions
will be addressed presently.

%}}}

\section{Results} %{{{
\label{sec:results}

It is of interest to examine whether cross-correlated or uncorrelated initial
conditions affect the long-term behavior of the system.  Two representative
numerical solutions are presented here that reveal the system's tendency to
form spatially-correlated structures in electron density and current regardless
of initial phase correlations.  This study establishes the robustness of
density structure formation in KAW turbulence and lends confidence that such
structures should exist in the ISM under varying circumstances.  The first
numerical solution has cross-correlated initial conditions between the $n$ and
$\psi$ fields; the second, uncorrelated.  Damping parameters $\eta$ and $\mu$
are equal and large enough to ensure numerical stability while preserving
structures in density, current and magnetic fields.  These examples also serve
to explore the intermediate $\eta / \mu$ regime.

\begin{figure}
\includegraphics[width=\textwidth]{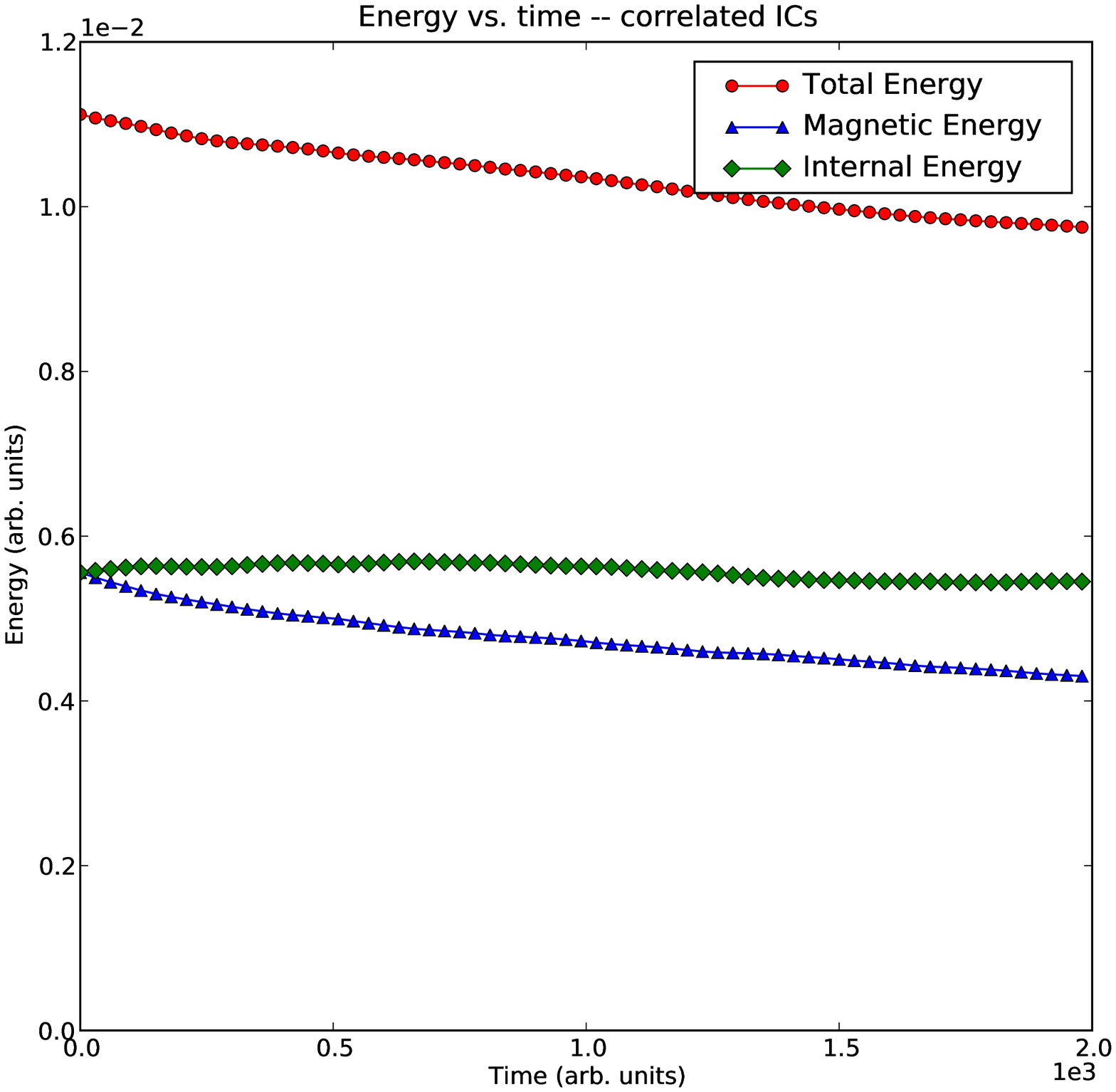}
\caption{Energy vs. time for cross-correlated initial conditions.  Total energy
is monotonically decreasing with time, and magnetic and internal energies
remain in rough equipartition.}
\label{energy-time-correlated}
\end{figure}

\begin{figure}
\includegraphics[width=\textwidth]{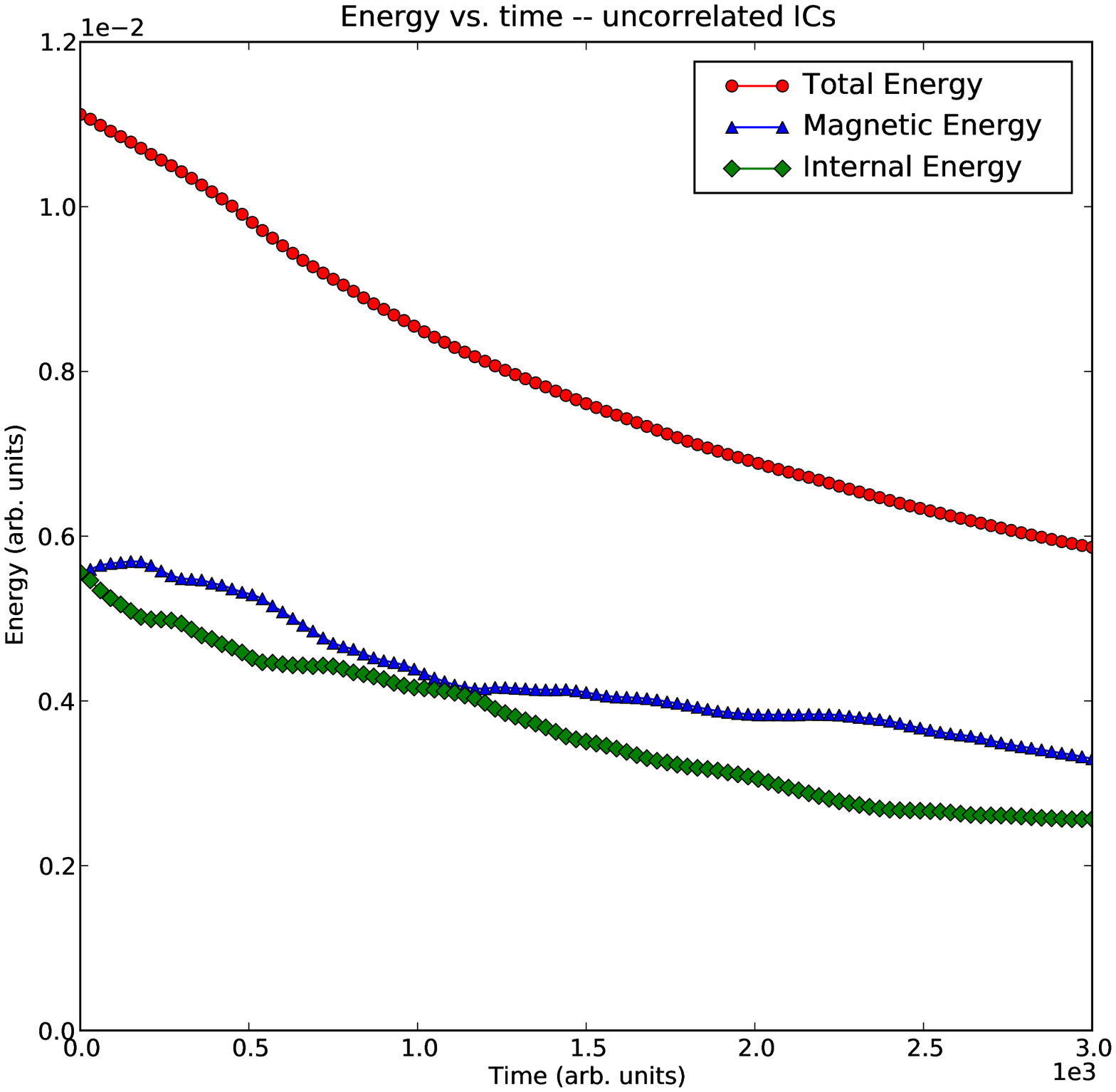}
\caption{Energy vs. time for uncorrelated initial conditions.  Total energy is
monotonically decreasing with time, and magnetic and internal energies remain
in rough equipartition.}
\label{energy-time-uncorrelated}
\end{figure}

The energy vs.\ time history for both numerical solutions are given in Figs.\
\ref{energy-time-correlated} and \ref{energy-time-uncorrelated}.  Total energy
is a monotonically decreasing function of time. The magnetic and internal
energies remain in overall equipartition throughout the numerical solutions.
Magnetic energy increases at the expense of internal energy and \emph{vice
versa}. This energy interchange is consistent with KAW dynamics and overall
energy conservation in the absence of resistive or diffusive terms.  The
exchange is crucial in routinely producing large amplitude density fluctuations
in this two-field model of nonlinearly interacting KAWs.

The total energy decay rates for the uncorrelated and correlated initial
conditions in Figs.\ \ref{energy-time-correlated} and
\ref{energy-time-uncorrelated} differ, with the latter decaying more strongly
than the former.  The damping parameters are identical for the two numerical
solutions, and the decay-rate difference remains under varying randomization
seeds.  The magnitudes of the nonlinear terms during the span of a numerical
solution in Eqs.\ (\ref{psi-eqn}) and (\ref{den-eqn}) for uncorrelated initial
conditions are consistently larger than those of correlated initial conditions
by a factor of 5. This difference lasts until 2500 Alfv\'en times, after which
the decay rates are roughly equal in magnitude.  The steeper energy decay
during the run of numerical solutions with uncorrelated initial conditions
(Fig.\ \ref{energy-time-uncorrelated}) suggests that the enhancement of the
uncorrelated nonlinearities transports energy to higher $k$ (smaller scale)
more readily than the nonlinearities in the correlated case.  Relatively more
energy in higher $k$ enhances the energy decay rate as the linear damping terms
dissipate more energy from the system.  The initial configuration, whether
correlated or uncorrelated, is seen to have an effect on the long-term energy
evolution for these decaying numerical solutions.  It will be shown below,
however, that the correlation does not significantly affect the statistics of
the resulting fields.

\begin{figure}
\includegraphics[width=\textwidth]{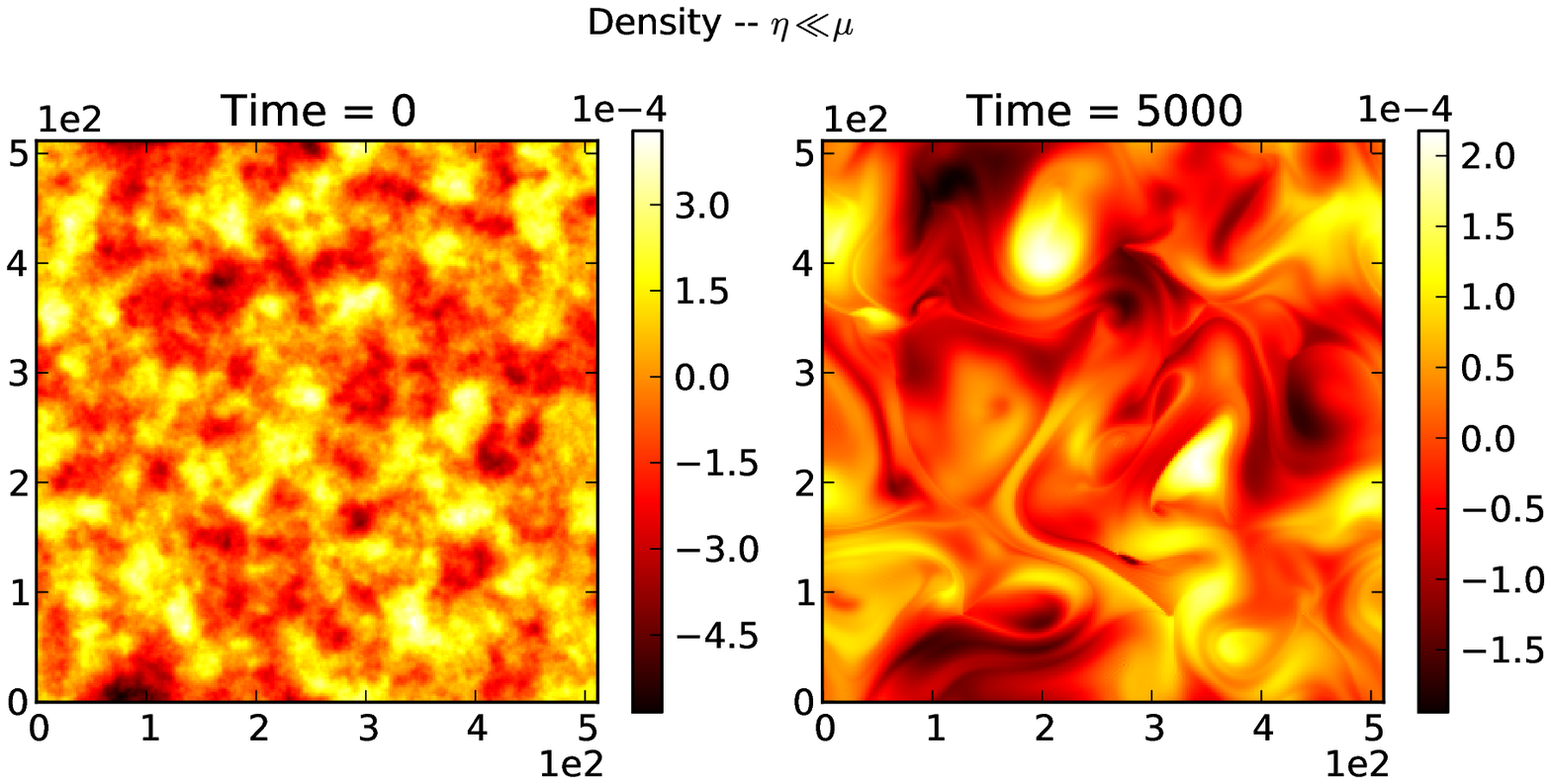}
\caption{Contours of $n$ for various times in a numerical solution with
correlated initial conditions.}
\label{dendat-correlated-contour}
\end{figure}

\begin{figure}
\includegraphics[width=\textwidth]{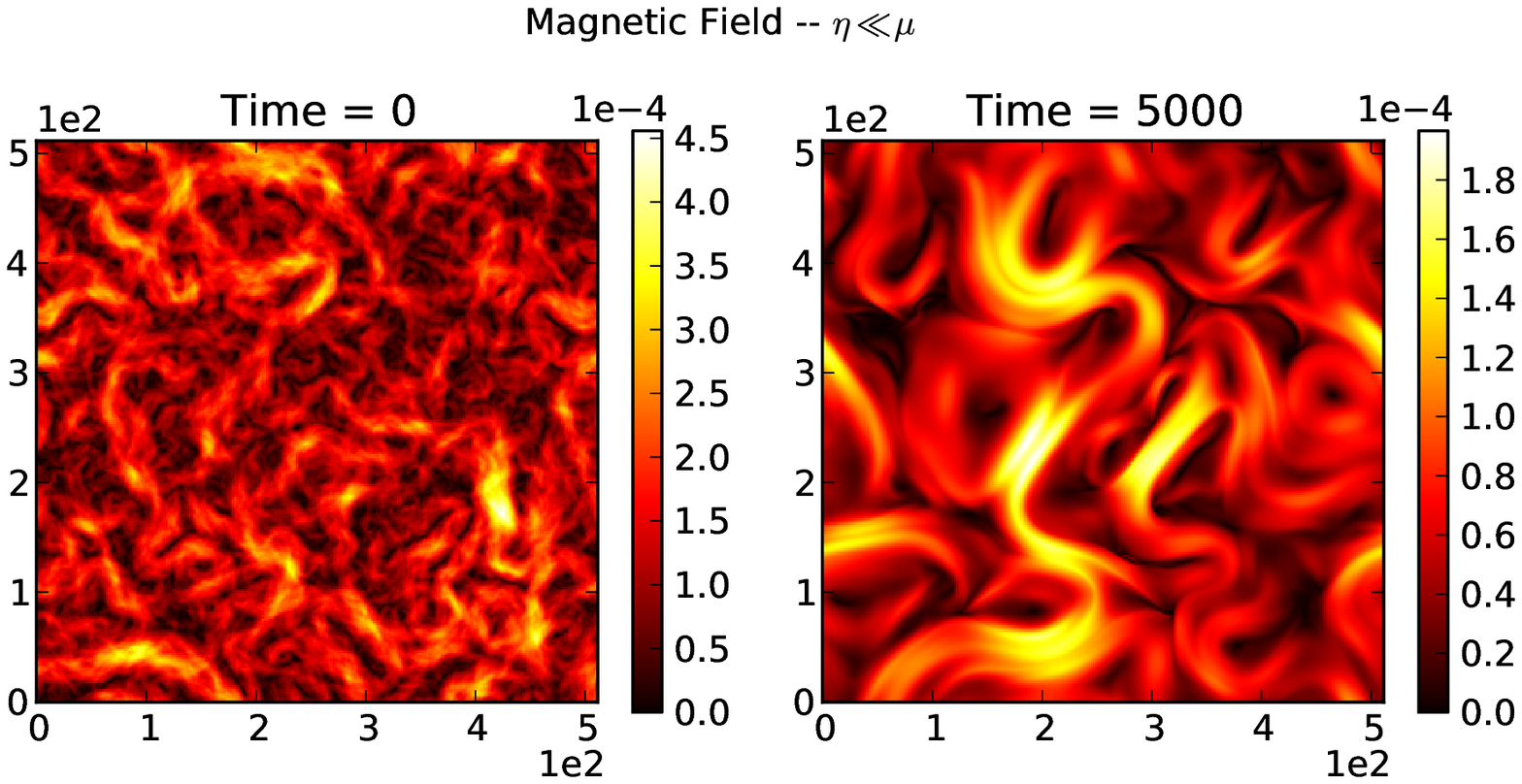}
\caption{Contours of $|\mathrm{B}|$ for various times in a numerical solution
with correlated initial conditions.}
\label{bmag-correlated-contour}
\end{figure}

\begin{figure}
\includegraphics[width=\textwidth]{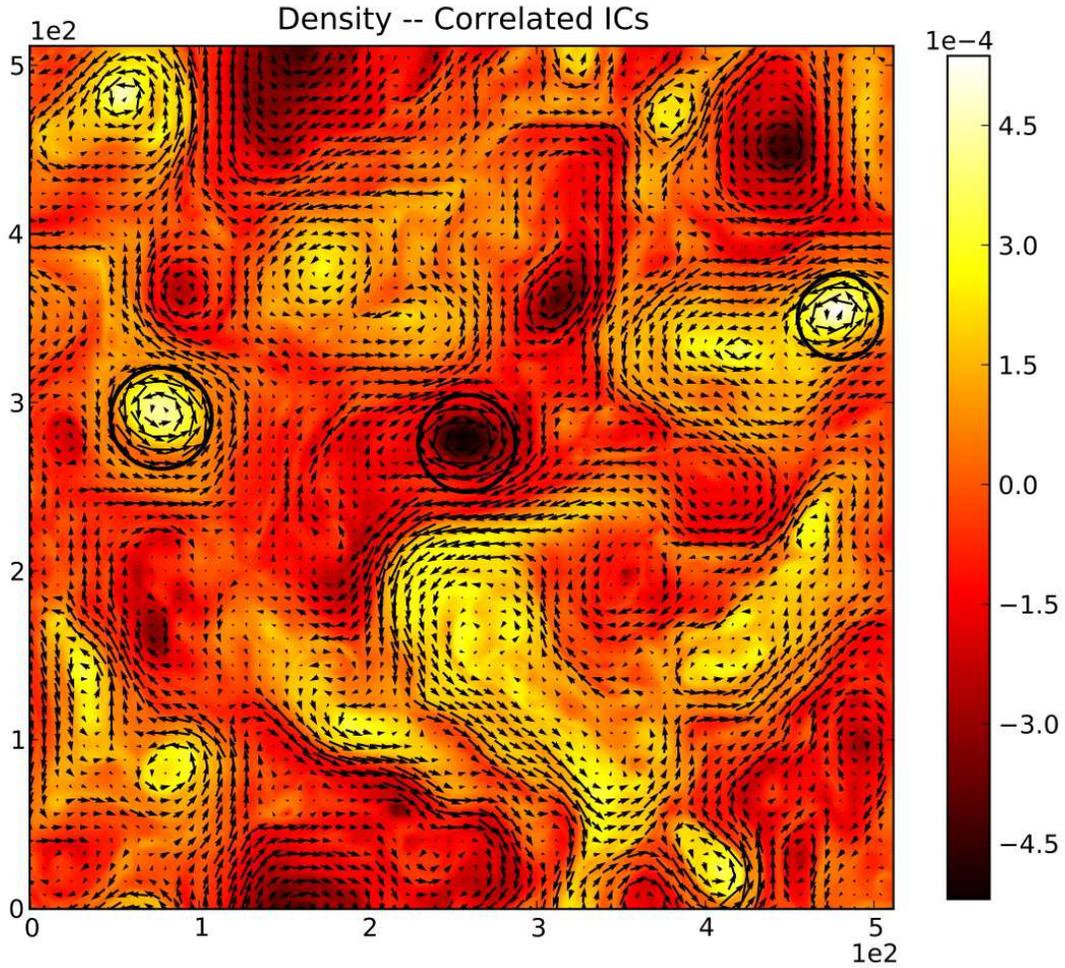}
\caption{Contour plot of $n$ with $\mathbf{B}$ vectors overlaid. The
positive, circularly-symmetric density structures correspond to
counterclockwise-directed $\mathbf{B}$ structures; the opposite holds for
negative circularly-symmetric density structures.  These spatial correlations
are to be expected for correlated initial conditions.}
\label{density-quiver-correlated}
\end{figure}

For cross-correlated initial conditions, we expect there to be a strong spatial
relation between current, magnetic field and density structures through time.
Figs.\ \ref{dendat-correlated-contour} and \ref{bmag-correlated-contour} show
the $n$ and $|\mathrm{B}|$ contours at various times.  For the latest time
contour, the spatial structure alignment is evident.  Further, in Fig.\
\ref{density-quiver-correlated}, the circular magnetic field structures
(magnetic field direction and intensity indicated by arrow overlays) align with
the large-amplitude density fluctuations.  The correlation is evident once one
notices that every positive-valued circular $n$ structure corresponds to
counterclockwise-oriented magnetic field, and \emph{vice versa}.  Fig.\
\ref{density-quiver-correlated} is at a normalized time of 5000 Alfv\'en times,
defined in terms of the large $\mathrm{B}_0$.  The system preserves the spatial
structure correlation indefinitely, even after structure mergers.

\begin{figure}
\includegraphics[width=\textwidth]{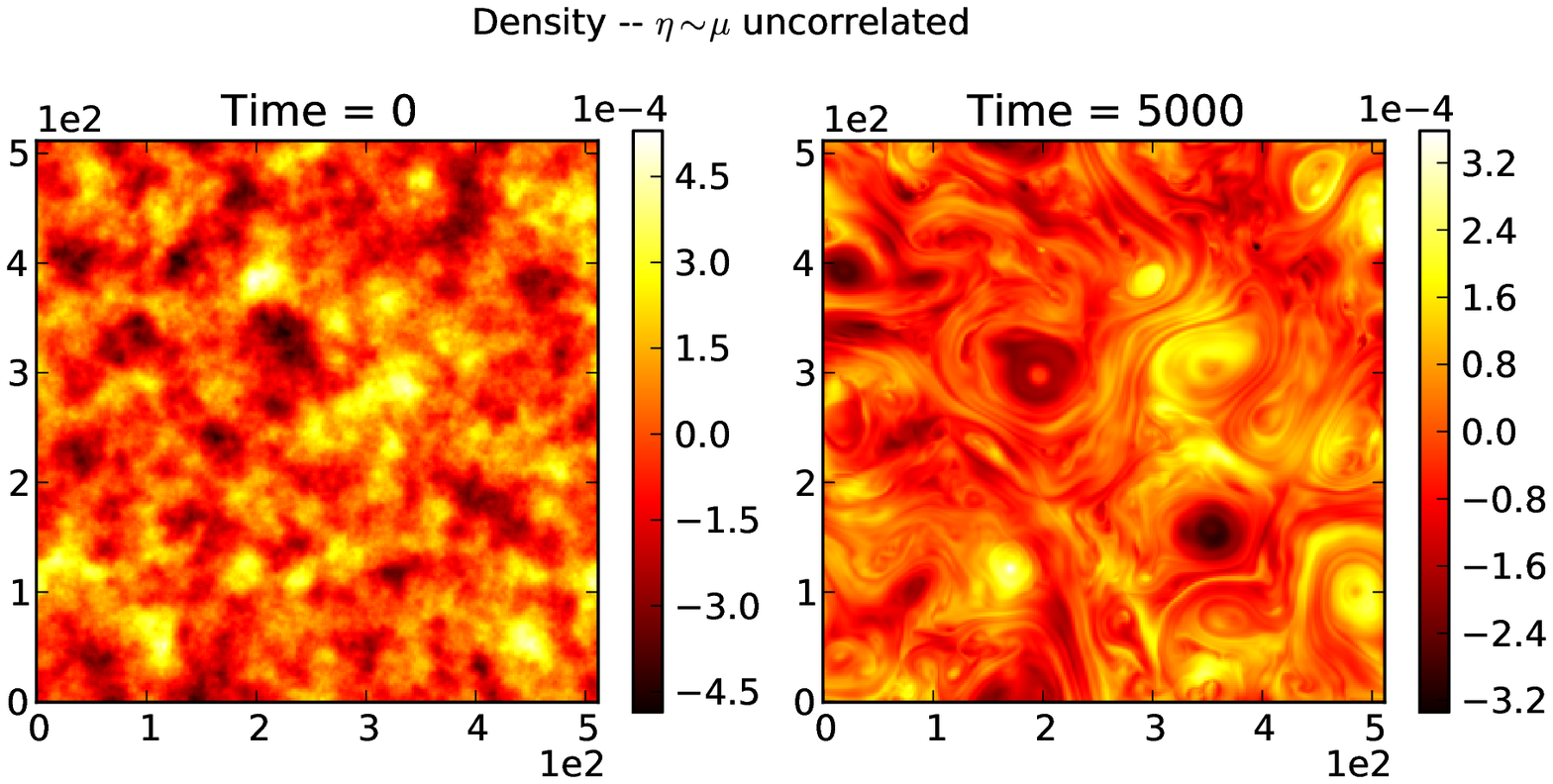}
\caption{Contours of $n$ for various times in a numerical solution with
uncorrelated initial conditions.}
\label{dendat-uncorrelated-contour}
\end{figure}

\begin{figure}
\includegraphics[width=\textwidth]{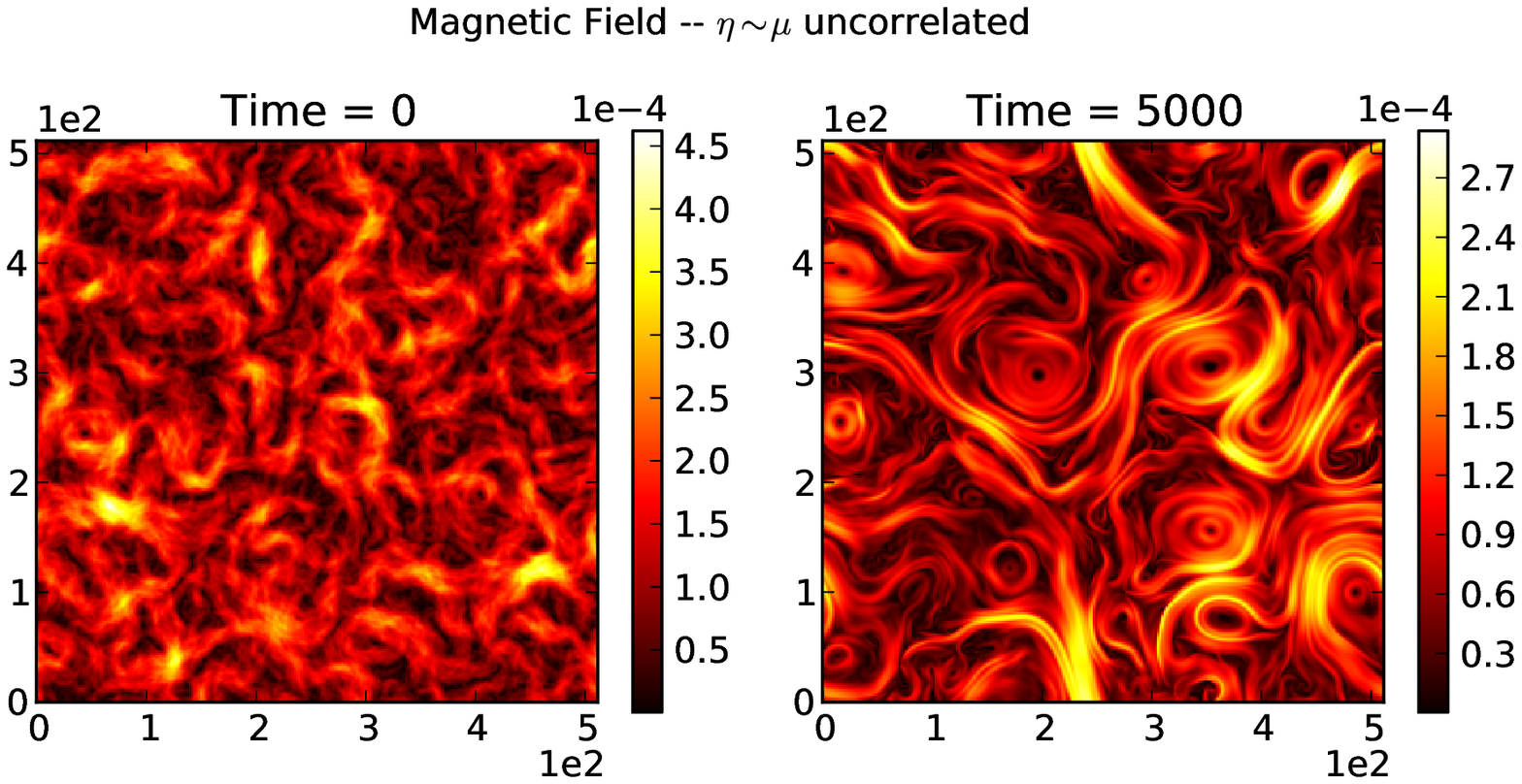}
\caption{Contours of $|\mathrm{B}|$ for various times in a numerical solution
with uncorrelated initial conditions.}
\label{bmag-uncorrelated-contour}
\end{figure}

The second representative numerical solution is one with uncorrelated initial
conditions.  Contour plots of density and $|\mathrm{B}|$ are given in Figs.\
\ref{dendat-uncorrelated-contour} and \ref{bmag-uncorrelated-contour},
respectively.  It is noteworthy that, similar to the cross-correlated initial
conditions, spatially correlated density and magnetic field structures are
discernible at the latest time contour.

\begin{figure}
\includegraphics[width=\textwidth]{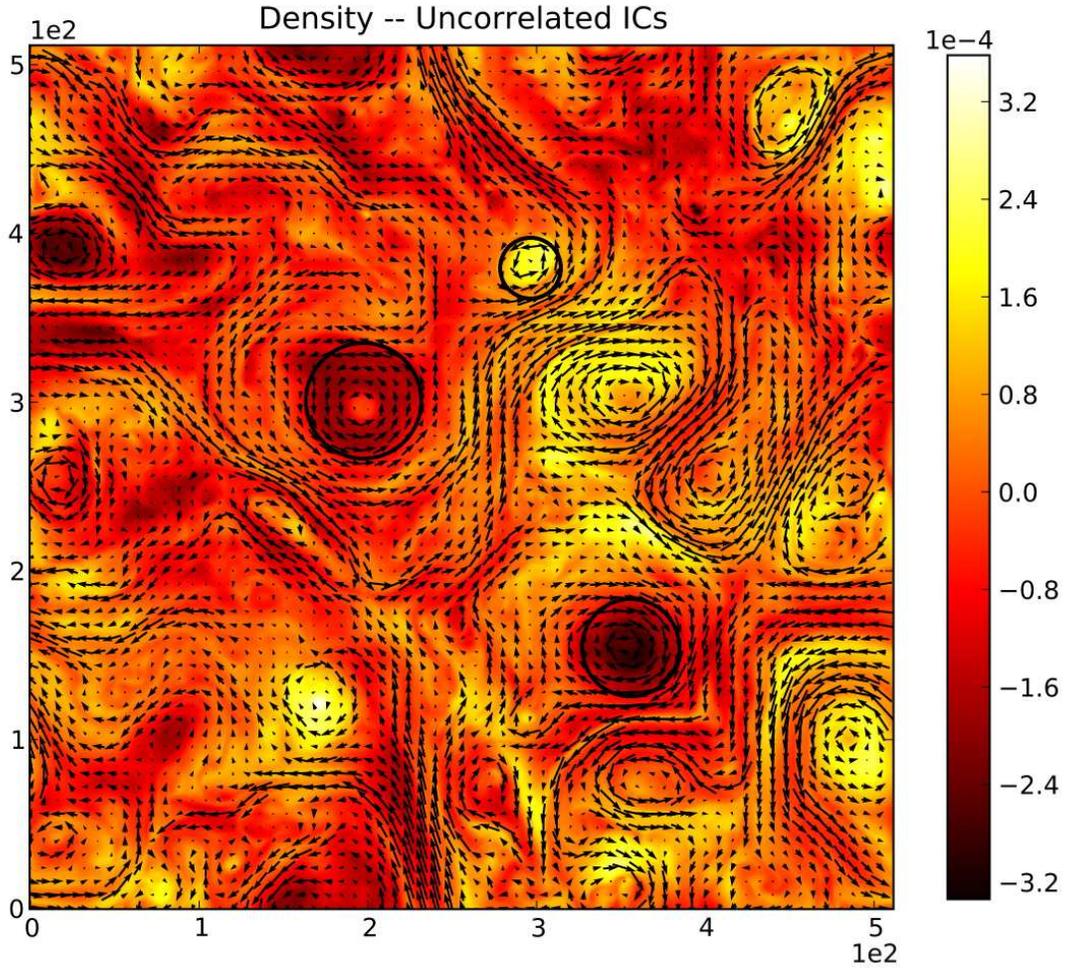}
\caption{Contour plot of $n$ with $\mathbf{B}$ vectors overlaid for a numerical
solution with initially uncorrelated initial conditions.  The positive,
circularly-symmetric density structures correspond to magnetic field
structures, although the sense (clockwise or counterclockwise) of the magnetic
field structure does not correlate with the sign of the density structures.
Circled in black are symmetric structures that display a high degree of spatial
correlation.  The circle gives an approximate indication of the separatrix for
the structure.}
\label{density-quiver-uncorrelated}
\end{figure}

In Fig.\ \ref{density-quiver-uncorrelated} the circular density structures
correspond to circular magnetic structures. Unlike Fig.\
\ref{density-quiver-correlated} the positive density structures may correspond
to clockwise or counterclockwise directed magnetic field structures.  This
serves to illustrate that, although the initial conditions have no phase
relation between fields, after many Alfv\'en times circular density structures
spatially correlate with magnetic field structures and persist for later times.

\begin{figure}
\includegraphics[width=\textwidth]{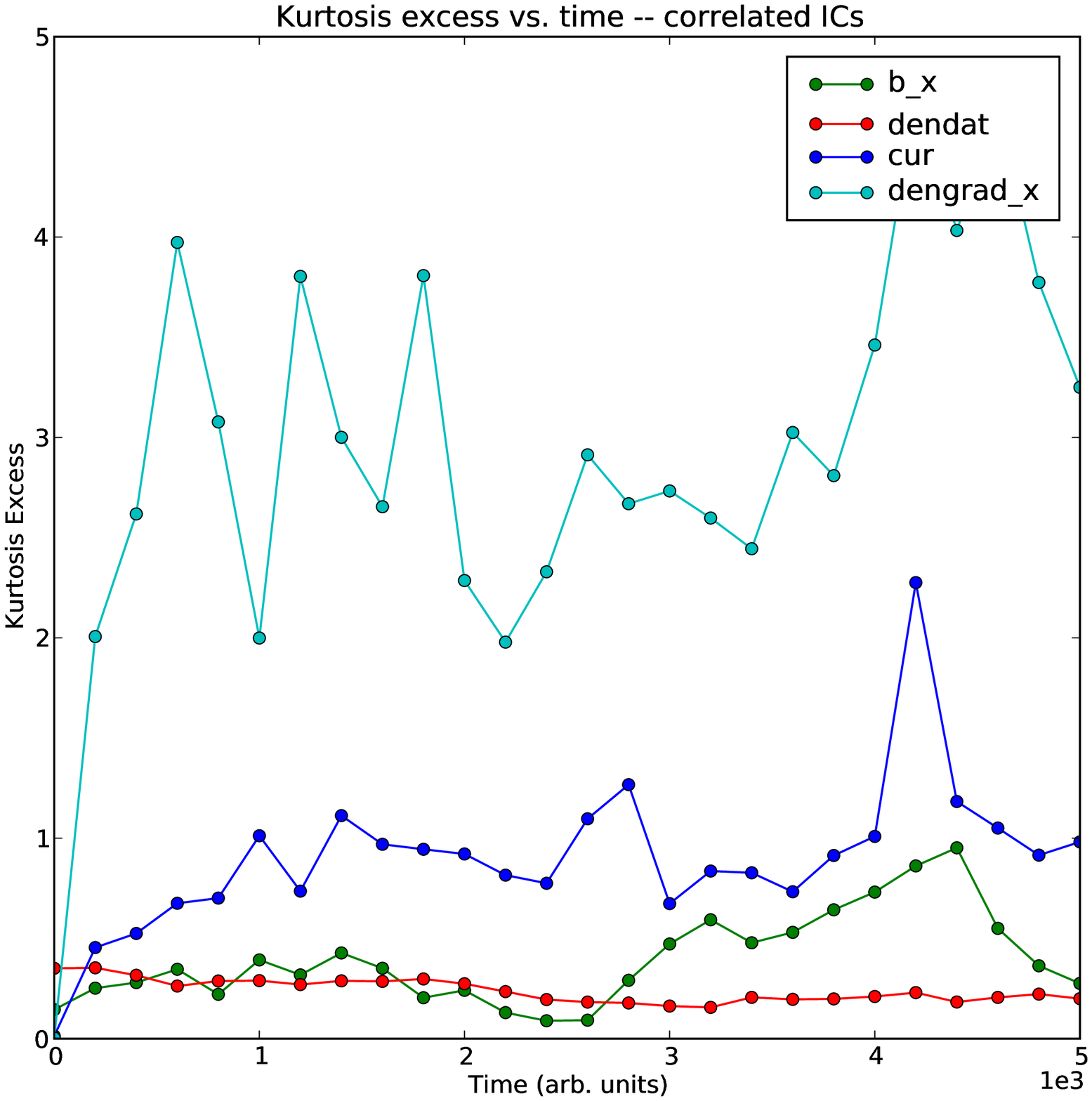}
\caption{Kurtosis excess for a numerical solution with phase-correlated initial
conditions and $\eta / \mu = 1$.}
\label{kurtosis-correlated}
\end{figure}

\begin{figure}
\includegraphics[width=\textwidth]{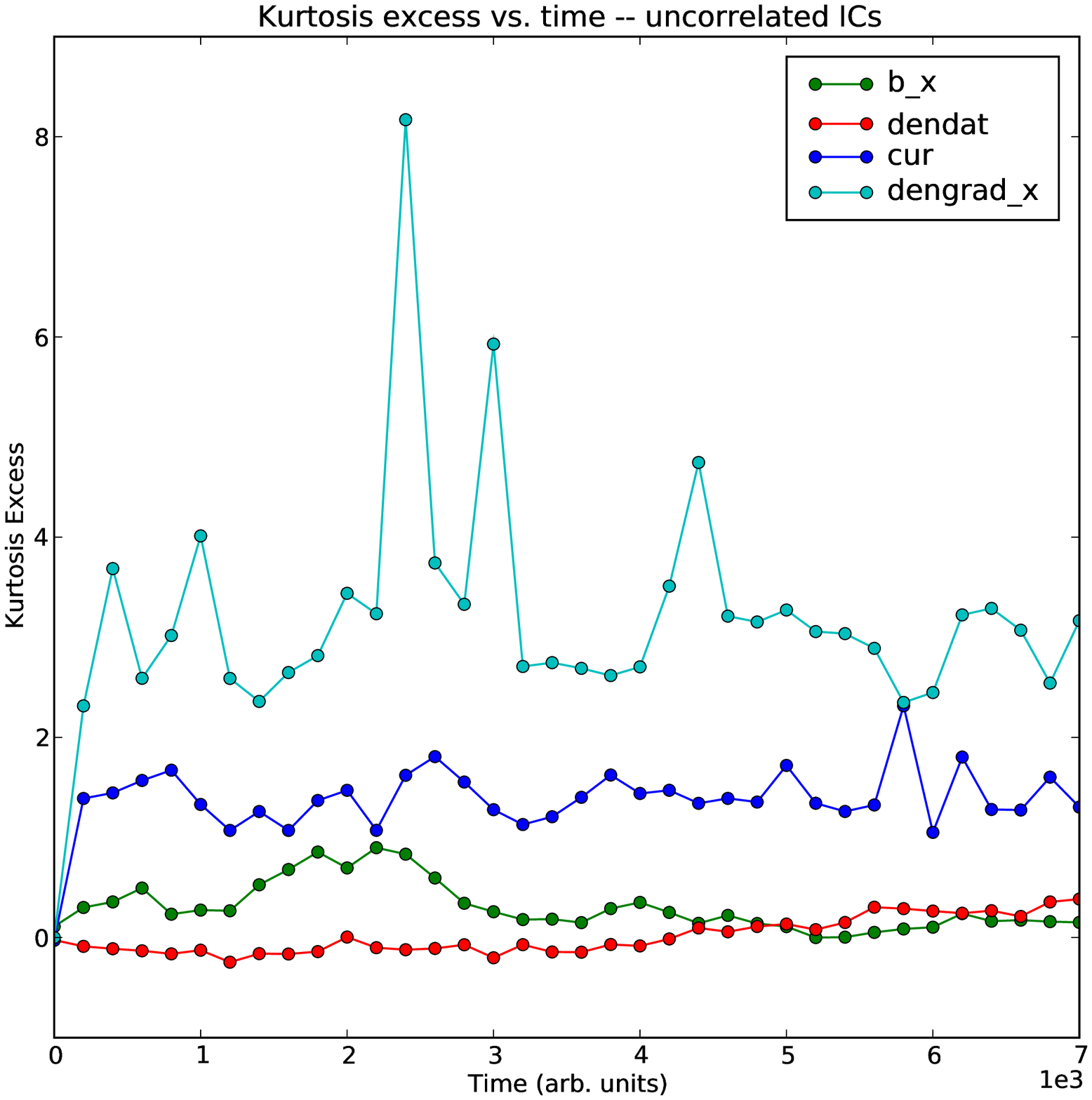}
\caption{Kurtosis excess for a numerical solution with phase-uncorrelated initial
conditions and $\eta / \mu = 1$.}
\label{kurtosis-uncorrelated}
\end{figure}

The kurtosis excess as a function of time, defined as
$K(\Xi)=\left<\Xi^4\right>/\sigma_{\Xi}^4 - 3$, is shown in Figs.\
\ref{kurtosis-correlated} and \ref{kurtosis-uncorrelated} for correlated and
uncorrelated initial conditions, respectively.  Positive $K$ indicates a
greater fraction of the distribution is in the tails as compared to a best-fit
Gaussian.  These figures indicate that the non-Gaussian statistics for the
fields of interest are independent of initial correlation in the fields.  In
particular, the density gradients, $|\nabla n|$, are significantly non-Gaussian
as compared to the current.  Because scintillation is tied to density
gradients, this situation is expected to favor the scaling inferred from pulsar
signals.

The tendency of density structures to align with magnetic field structures
regardless of initial conditions indicates that the initial conditions are
representative of fully-developed turbulence. After a small number of Alfv\'en
times the memory of the initial state is removed as the KAW interaction sets up
a consistent phase relation between the fluctuations in the magnetic and
density fields.  Previous work \citep{terry-smith07} presented a mechanism
whereby these spatially correlated structures can be preserved via shear in the
periphery of the structures. The above figures indicate that this mechanism is
at play even in cases where the initial phase relations are uncorrelated.

In the damping regime presented above, circularly symmetric structures in
density, current and magnetic fields readily form and persist for many Alfv\'en
times, until disrupted by mergers with other structures of similar amplitude.
It is possible to define, for each circular structure, an effective separatrix
that distinguishes it from surrounding turbulence and large-amplitude
``sheets'' that exist between structures. [see, e.g., the magnetic field
contours at later times in Fig.\ \ref{bmag-uncorrelated-contour}.]  The density
field has significant gradients in both the regions surrounding the structure
and within the structures themselves.  The ability to separate these circular
structures from the background sheets and turbulence is determined by the
magnitudes -- relative and absolute -- of the damping parameters.  Larger
damping values erode the small-spatial-scale structures to a greater extent
and, if large enough, disrupt the structure persistence mechanism that, for a
fixed diameter, depends on a sufficiently large amplitude current filament to
generate a sufficiently large radially sheared magnetic field.

The preceding results were for a damping regime where $\eta / \mu \sim 1$, an
intermediate regime.  Numerical solutions with $\mu = 0$ and $\eta$ small
explore the regime where $\eta / \mu \rightarrow 0$.  In this regime, which is
opposite the regime used in Craddock et al., circularly symmetric current and
magnetic structures are not as prevalent, rather, sheet-like structures
dominate the large amplitude fluctuations.  Current and magnetic field
gradients are strongly damped, and the characteristic length scales in these
fields are larger.

\begin{figure}
\includegraphics[width=\textwidth]{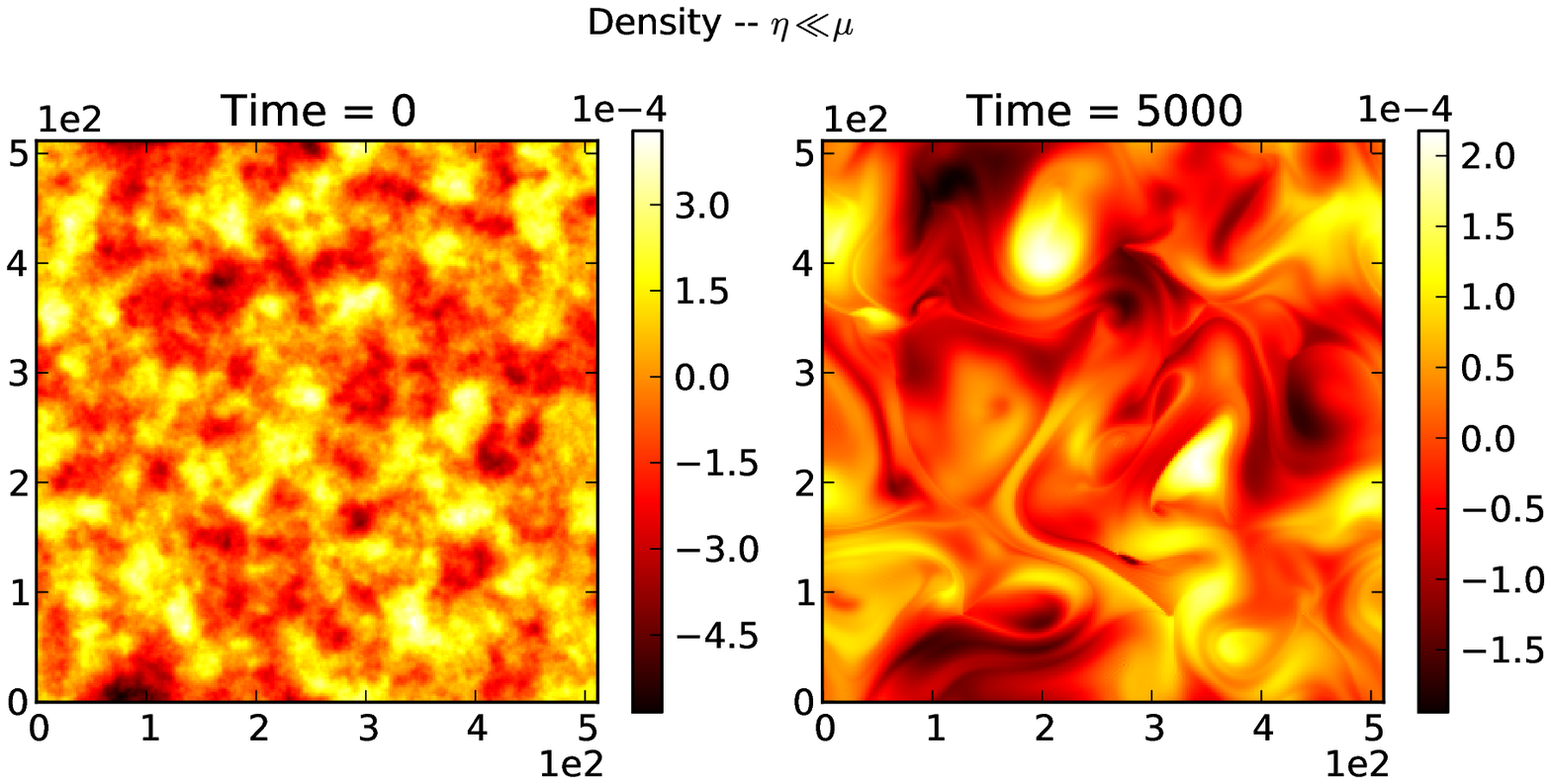}
\caption{Electron density contour visualization with diffusive damping
parameter $\mu = 0$ for various times.}
\label{fig:density-zero-mu}
\end{figure}

\begin{figure}
\includegraphics[width=\textwidth]{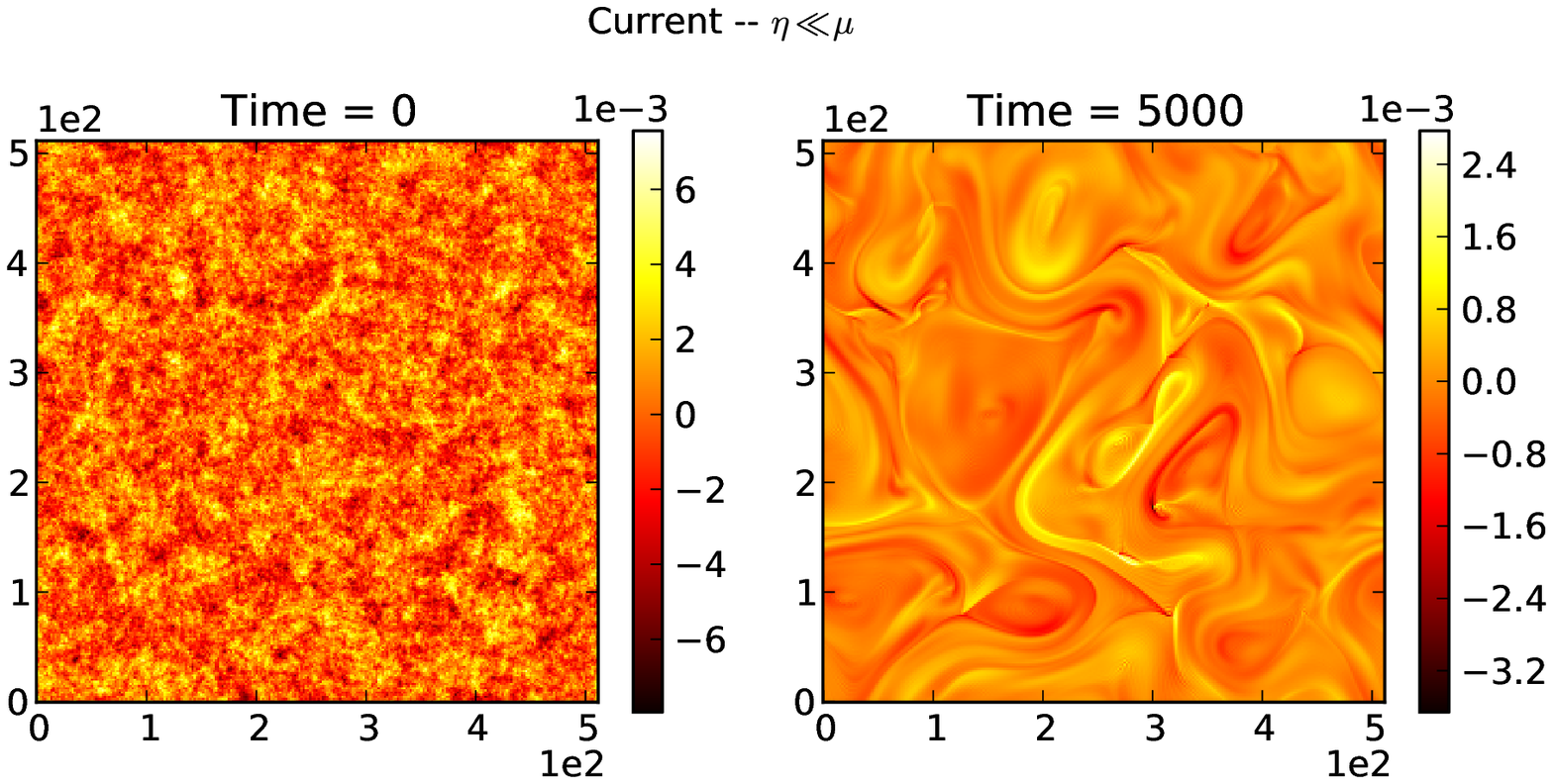}
\caption{Current density contour visualization with diffusive damping parameter
$\mu = 0$ for various times.}
\label{fig:current-zero-mu}
\end{figure}

\begin{figure}
\includegraphics[width=\textwidth]{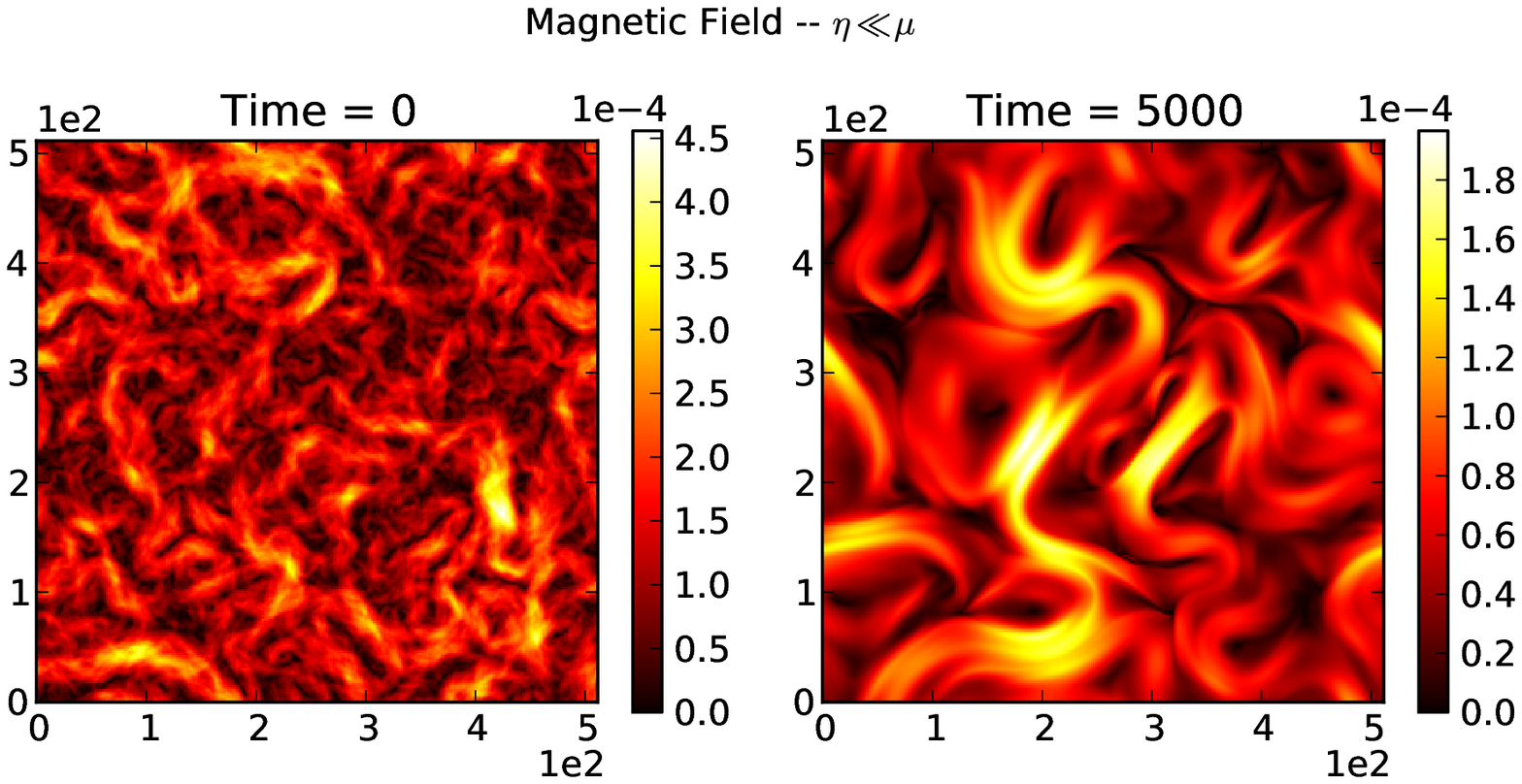}
\caption{Magnitude of magnetic field contour visualization with diffusive
damping parameter $\mu = 0$ for various times.} 
\label{fig:bmag-zero-mu}
\end{figure}

Contours of density for a numerical solution with $\mu = 0$ are shown in Fig.\
\ref{fig:density-zero-mu}.  Visual comparison with contours for runs with
smaller damping parameters (Fig.\ \ref{dendat-uncorrelated-contour}, where
$\eta = \mu$) indicate a preponderance of sheets in the $\mu = 0$ case, at the
expense of circularly-symmetric structures as seen above.  All damping is in
$\eta$; any current filament that would otherwise form is unable to preserve
its small-scale, large amplitude characteristics before being resistively
damped.  Inspection of the current and $|B|$ contours for the same numerical
solution [Figs.\ \ref{fig:current-zero-mu} and \ref{fig:bmag-zero-mu}] reveal
broader profiles and relatively few circular current and magnetic field
structures with a well-defined separatrix as in the small $\eta$ case.  Since
there is no diffusive damping, gradients in electron density are able to
persist, and electron density structures generally follow the same structures
in the current and magnetic fields.

\begin{figure}
\includegraphics[width=\textwidth]{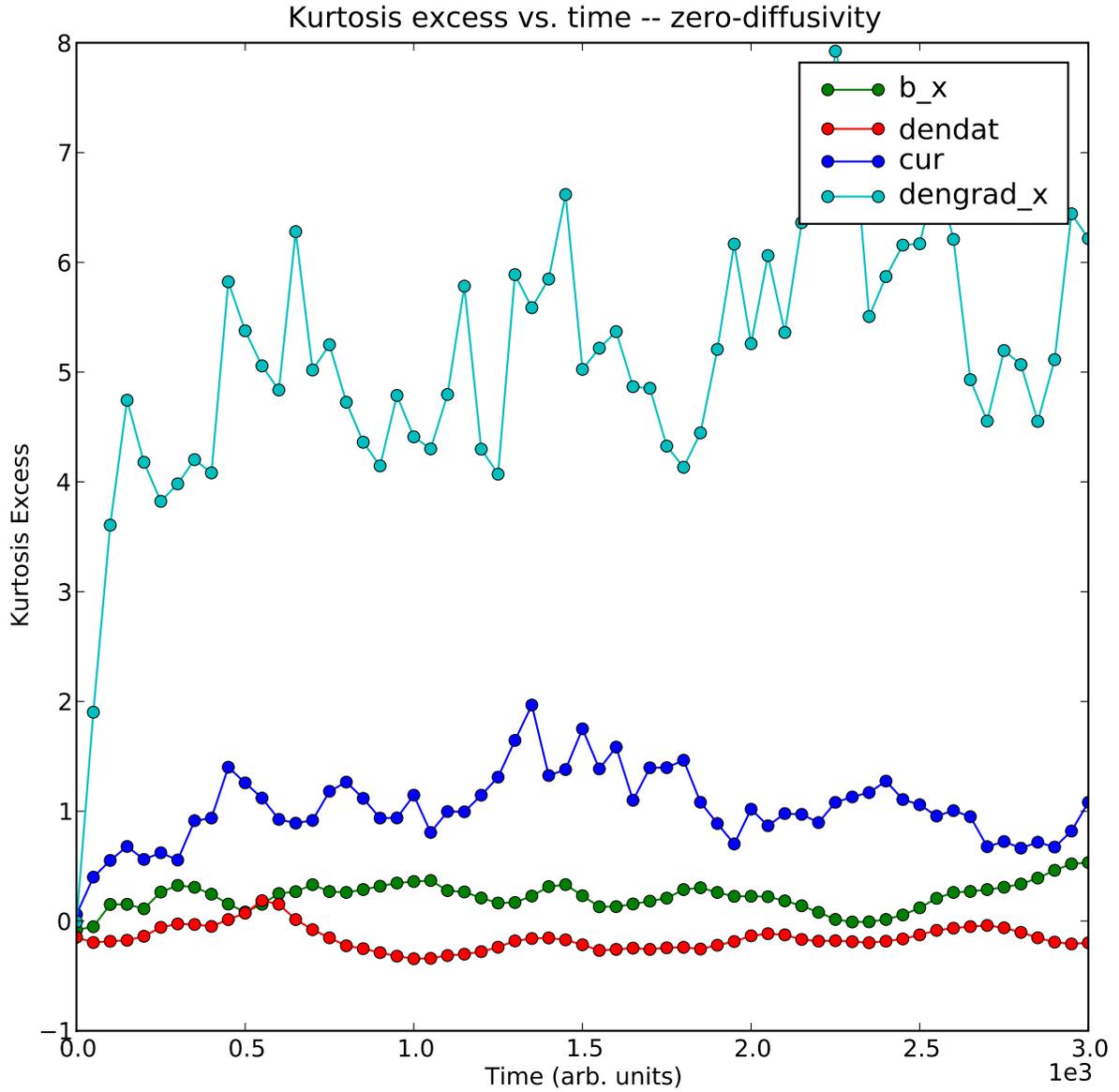}
\caption{Kurtosis excess for a numerical solution with diffusive parameter $\mu
= 0$.  Density gradient kurtosis remains greater than current kurtosis for the
duration of the numerical solution.}
\label{fig:zero-mu-kurtosis}
\end{figure}

\begin{figure}
\includegraphics[width=\textwidth]{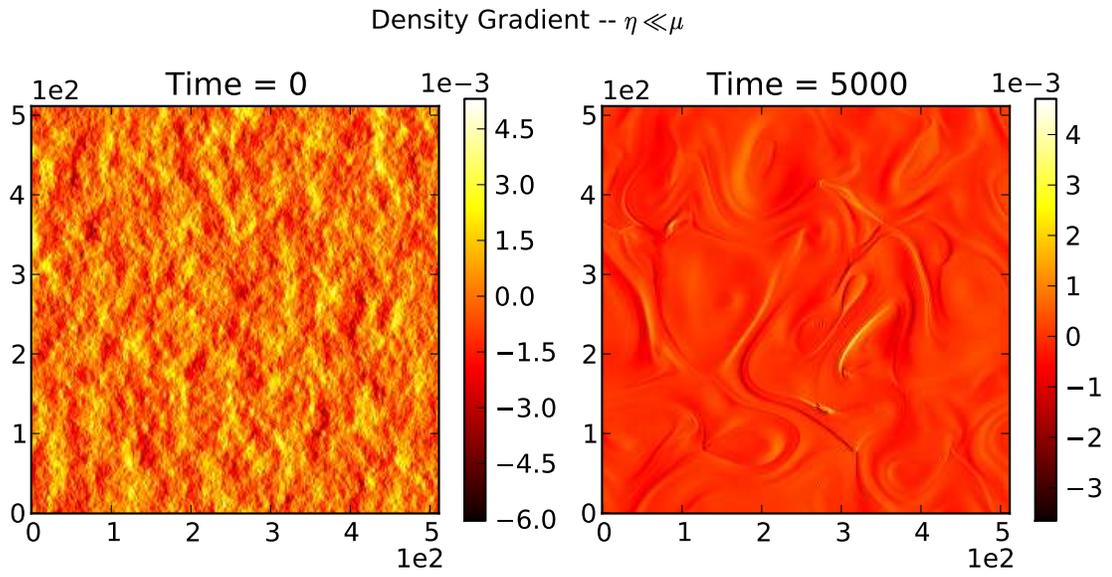}
\caption{Electron density gradient ($x$ direction) contour visualization with
diffusive damping $\mu = 0$ for various times.}
\label{fig:dengrad-zero-mu}
\end{figure}

Kurtosis excess measurements for the $\mu = 0$ numerical solutions yield mean
values consistent with the $\eta = \mu$ numerical solutions, as seen in Fig.\
\ref{fig:zero-mu-kurtosis}.  Magnetic field strength and electron density
statistics are predominantly Gaussian, with current statistics and density
gradient statistics each non-Gaussian.  Perhaps not as remarkable in this case,
the density gradient kurtosis excess is again seen to be greater than the
current kurtosis excess -- this is anticipated since the dominant damping of
density gradients is turned off.  With fewer filamentary current structures,
however, the mechanism proposed in \citet{terry-smith07} is not likely to be at
play in this case, since few large-amplitude filamentary current structures
exist.  Sheets, evident in the density gradients in Fig.\
\ref{fig:dengrad-zero-mu} and in the current in Fig.\ \ref{fig:current-zero-mu}
are the dominant large-amplitude structures and determine the extent to which
the density gradients have non-Gaussian statistics.  The current and density
sheets are well correlated spatially.  The largest sheets can extend through
the entire domain, and evolve on a longer timescale than the turbulence.
Sheets exist at the interface between large-scale flux tubes, and are regions
of large magnetic shear, giving rise to reconnection events.  With $\eta$
relatively large, the sheets evolve on timescales shorter than the structure
persistence timescale associated with the long-lived flux tubes.

Sheets and filaments are the dominant large-amplitude, long timescale
structures that arise in the KAW system.  Filaments arise and persist as long
as $\eta$ is small, with their amplitude and statistical influence diminished
as $\eta$ increases.  Sheets exist in both regimes, becoming the sole
large-scale structure in the large $\eta$ regime.  Density gradients are
consistently non-Gaussian in both regimes as long as $\mu$ is small, although
the density structures are different in both regimes.  Density gradient sheets
arise in the large $\eta$ regime and these density gradient sheets are large
enough to yield non-Gaussian statistics.

%}}}

\section{Ensemble Statistics and PDFs}%{{{
\label{sec:PDFs}

To explicitly analyze the extent to which the decaying KAW system develops
non-Gaussian statistics, ensemble runs were performed for both the $\eta / \mu
\sim 1$ and $\eta / \mu \ll 1$ regimes, and PDFs of the fields were generated.

\begin{figure}
\includegraphics[width=\textwidth]{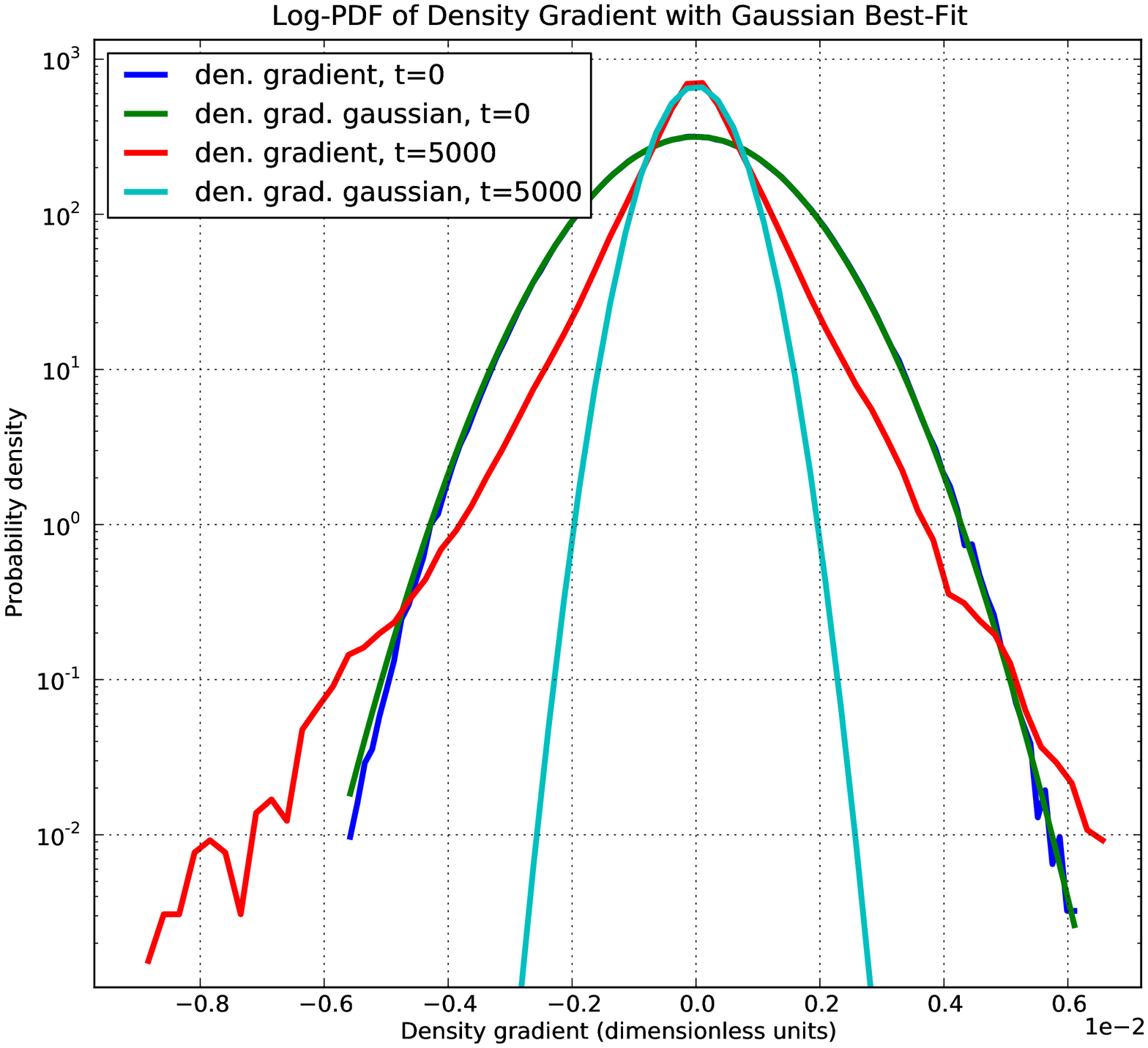}
\caption{log-PDF of density gradients for an ensemble of numerical solutions
with $\eta / \mu = 1$ at $t=0$ and $t=5000$.  The density gradient field at
$t=0$ is Gaussian distributed, while for $t=5000$ the gradients are enhanced in
the tails, and deviate from a Gaussian.  A best-fit Gaussian for each PDF is
plotted for comparison.}
\label{fig:dgx-PDF}
\end{figure}

\begin{figure}
\includegraphics[width=\textwidth]{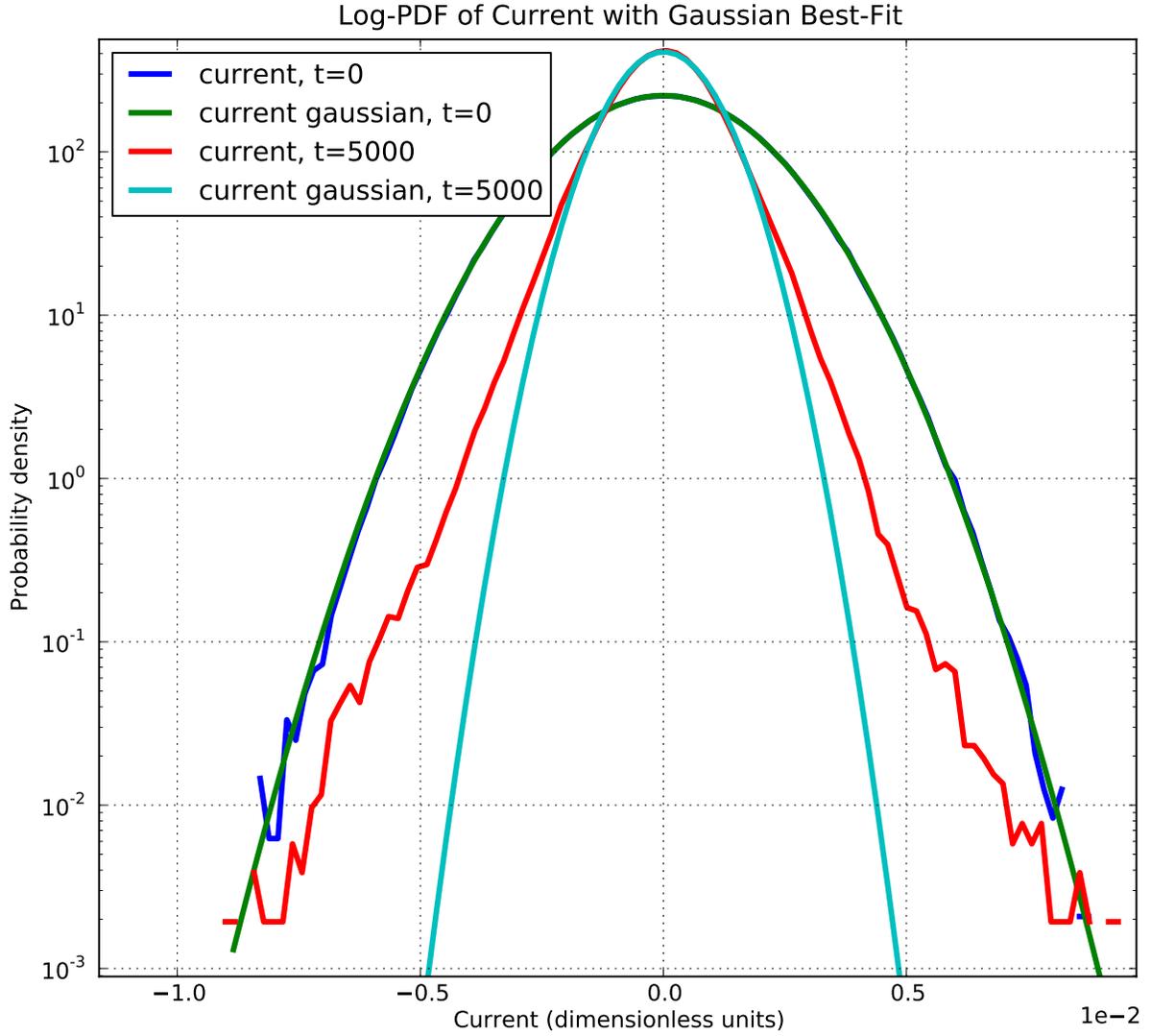}
\caption{log-PDF of current for an ensemble of numerical solutions with $\eta /
\mu = 1$ at $t=0$ and $t=5000$.  The current at $t=0$ is Gaussian distributed.
For $t=5000$ the current is non-Gaussian.  Unlike the density gradient, the
current is not enhanced in the tails of the PDF for later times relative to its
initial Gaussian envelope.}
\label{fig:cur-PDF}
\end{figure}

For the $\eta / \mu \sim 1$ regime, 10 numerical solutions were evolved with
identical parameters but for different randomization seeds.  In this case $\eta
= \mu$ and both damping parameters have minimal values to ensure numerical
stability.  The fields were initially phase-uncorrelated.  The density gradient
ensemble PDF for two times in the solution results is shown in Fig.\
\ref{fig:dgx-PDF}. Density gradients are Gaussian distributed initially.  Many
Alfv\'en times into the numerical solution the statistics are non-Gaussian with
long tails.  These PDFs are consistent with the time histories of density
gradient kurtosis excess as shown above.  The distribution tail extends beyond
15 standard deviations, almost 90 orders of magnitude above a Gaussian best-fit
distribution.  Similar behavior is seen in the current PDFs -- initially
Gaussian distributed tending to strongly non-Gaussian statistics with long
tails for later times.  Fig.\ \ref{fig:cur-PDF} is the current PDF at an
advanced time into the numerical solution.  It is to be noted that the density
gradient PDF has longer tails at higher amplitude than does the current PDF.
One would expect these to be in rough agreement, since the underlying density
and magnetic fields have comparable PDFs that remain Gaussian distributed
throughout the numerical solution.  The discrepancy between the density
gradient and current PDFs suggests a process that enhances density derivatives
above magnetic field derivatives.  Future work is required to explore causes of
this enhancement.  This result is significant for pulsar scintillation, which
is most sensitive to density gradients.  Although interstellar turbulence is
magnetic in nature, the KAW regime has the benefit of fluctuation equipartition
between $n$ and $B$.  The density gradient, however, is more non-Gaussian than
the magnetic component, suggesting that this type of turbulence is specially
endowed to produce the type of scintillation scaling observed with pulsar
signals.

\begin{figure}
\includegraphics[width=\textwidth]{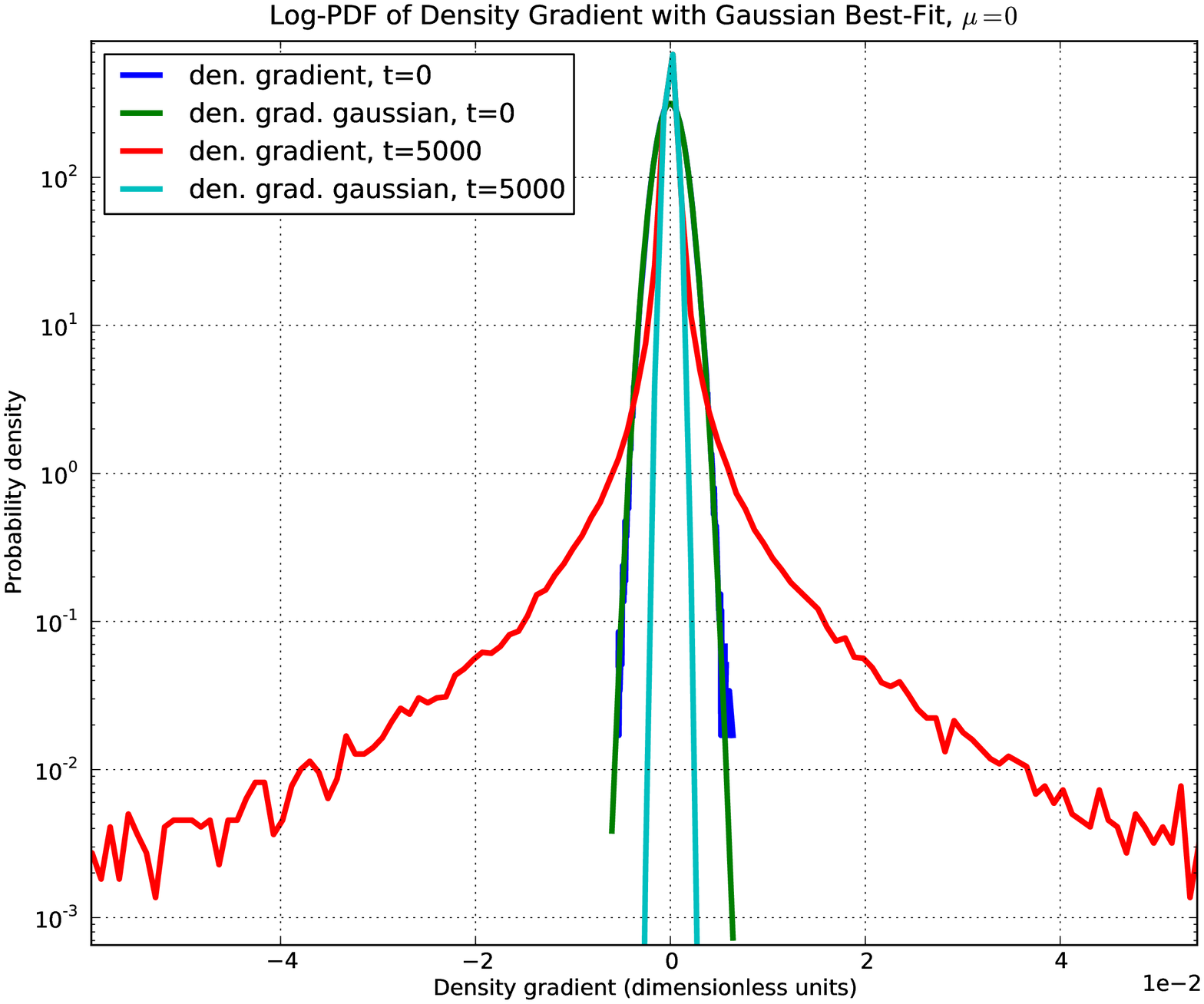}
\caption{Log-PDF of density gradient for an ensemble of numerical solutions
with $\mu=0$ at $t=0$ and $t=5000$.  The density gradient field at $t=0$ is
Gaussian distributed, while for $t=5000$ the gradients are enhanced in the
tails, and deviate from a Gaussian.  A best-fit Gaussian for each PDF is
plotted for comparison.}
\label{fig:dgx-PDF-zero-mu}
\end{figure}

Ensemble runs for the $\eta / \mu \ll 1$ regime yield distributions similar to
the $\eta / \mu \sim 1$ regime in all fields.  The ensemble PDF for two times
is shown in Fig.\ \ref{fig:dgx-PDF-zero-mu}. The initial density gradient PDF
is Gaussian distributed.  For later times long tails are evident and consistent
with the kurtosis excess measurements as presented above for the $\mu = 0$
case.  The density gradient distribution has longer tails at higher amplitude
than the current distribution; the overall distributions are similar to those
for the $\eta / \mu \sim 1$ regime, despite the absence of filamentary
structures and the presence of sheets.  The strongly non-Gaussian statistics
are insensitive to the damping regime, provided that the diffusion coefficient
is small enough to allow density gradients to persist.

%}}}

\section{Discussion}%{{{

Using the normalizations for Eqs.\ (\ref{psi-eqn}) and (\ref{den-eqn}) and
using $B=1.4 \mu$G, $n=0.08$ cm$^{-3}$ and $T_e=1$ eV, $\eta_{norm}$, the
normalized Spitzer resistivity, is $2.4 \times 10^{-7}$ and $\mu_{norm}$, the
normalized collisional diffusivity, is $1.9 \times 10^{-7}$.  For a resolution
of $512^2$, these damping values are unable to keep the system numerically
stable.  The threshold for stability requires the simulation $\eta$ to be
greater than $5 \times 10^{-6}$, which is almost within an order of magnitude
of the ISM value.  The numerical solutions presented here, while motivated by
the pulsar signal width scalings, more generally characterize the current and
density gradient PDFs when the damping parameters are varied.  We would expect
the density gradients to be non-Gaussian when using parameters that correspond
to the ISM.  Future work will address the pulsar width scaling using electron
density fields from the numerical solution.

The non-Gaussian distributions presented here are strongly tied to the fact
that the system is decaying and that circular intermittent structures are
preserved from nonlinear interaction.  One can show that, in the KAW system,
circularly symmetric structures (or filaments) are force free in Eqs.\
(\ref{psi-eqn}) and (\ref{den-eqn}), i.e., the nonlinearity is zero.  Once a
large-amplitude structure becomes sufficiently circularly symmetric and is able
to preserve itself from background turbulence via the shear mechanism, that
structure is expected to persist on long timescales relative to the turbulence.
Structure mergers will lead to a time-asymptotic state with two
oppositely-signed current structures and no turbulence.  As structures merge,
kurtosis excess increases until the system reaches a final two-filament state,
which would have a strongly non-Gaussian distribution and large kurtosis
excess.

If the system were driven, energy input at large scales would replenish
large-amplitude fluctuations.  New structures would arise from large amplitude
regions whenever the radial magnetic field shear were large enough to preserve
the structure from interaction with turbulence.  One could define a
structure-replenishing rate from the driving terms that would depend on the
energy injection rate and scale of injection.  The non-Gaussian measures for a
driven system would be characterized by a competition between the creation of
new structures through the injection of energy at large scales and the
annihilation of structures by mergers or by erosion from continuously
replenished small-scale turbulence.  If erosion effects dominate, the kurtosis
excess is maintained at Gaussian values, diminishing the PDF tails relative to
a L\'evy distribution.  If replenishing effects dominate, however, the
enhancement of the tails of the density gradient PDF may be observed in a
driven system as it is observed in the present decaying system.  We note that
structure function scaling in hydrodynamic turbulence is consistent with the
replenishing effects becoming more dominant relative to erosion effects as
scales become smaller, i.e., the turbulence is more intermittent at smaller
scales.  The large range of scales in interstellar turbulence and the
conversion of MHD fluctuations to kinetic Alfv\'en fluctuations at small scales
both support the notion that the structures of the decaying system are relevant
to interstellar turbulence at the scales of KAW excitations.  This scenario is
consistent with arguments suggested by \citet{harmon05}.  They propose a
turbulent cascade in the solar wind that injects energy into the KAW regime,
counteracting Landau damping at scales near the ion Larmor radius.  By doing so
they can account for enhanced small-scale density fluctuations and observed
scintillation effects in interplanetary scintillation.  

We also observe that, although the numerical solutions presented here are
decaying in time, the decay rate decreases in absolute value for later times
(Figs.\ \ref{energy-time-correlated} and \ref{energy-time-uncorrelated}),
approximating a steady-state configuration.  The kurtosis excess (Figs.\
\ref{kurtosis-correlated} and \ref{kurtosis-uncorrelated}) for the density
gradient field is statistically stationary after a brief startup period.
Despite the decaying character of the numerical solutions, they suggest that
the density gradient field would be non-Gaussian in the driven case.

The kurtosis excess -- a measure of a field's spatial intermittency -- is
itself intermittent in time.  The large spikes in kurtosis excess correspond to
rare events involving the merger of two large-amplitude structures, usually
filaments.  A large-amplitude short-lived sheet grows between the structures
and persists throughout the merger, gaining amplitude in time until the point
of merger.  The kurtosis excess during this merger event is dominated by the
single large-amplitude sheet between the merging structures.  This would likely
be the region of dominant scattering for scintillation, since a corresponding
large-amplitude density gradient structure exists in this region as well.  The
temporal intermittency of kurtosis excess suggests that these mergers are rare
and hence, of low probability.  The heuristic picture of long undeviated L\'evy
flights punctuated by large angular deviations could apply to these merger
sheets.

%}}}

\section{Conclusions} %{{{

Decaying kinetic Alfv\'en wave turbulence is shown to yield non-Gaussian
electron density gradients, consistent with non-Gaussian distributed density
gradients inferred from pulsar width scaling with distance to source.  With
small resistivity, large-amplitude current filaments form spontaneously from
Gaussian initial conditions, and these filaments are spatially correlated with
stable electron density structures.  The electron density field, while Gaussian
throughout the numerical solution, has gradients that are strongly
non-Gaussian.  Ensemble statistics for current and density gradient fields
confirm the kurtosis measurements for individual runs.  Density gradient
statistics, when compared to current statistics, have more enhanced tails, even
though both these fields are a single derivative away from electron density and
magnetic field, respectively, which are in equipartition and Gaussian
distributed throughout the numerical solution.  

When all damping is placed in resistive diffusion ($\eta / \mu \rightarrow 0$
regime), filamentary structures give way to sheet-like structures in current,
magnetic, electron density and density gradient fields.  Kurtosis measurements
remain similar to those for the small $\eta$ case, and the field PDFs also
remain largely unchanged, despite the different large-amplitude structures at
play.

The kind of structures that emerge, whether filaments or sheets, is a function
of the damping parameters.  With $\eta$ and $\mu$ minimal to preserve numerical
stability and of comparable value, the decaying KAW system tends to form
filamentary current structures with associated larger-scale magnetic and
density structures, all generally circularly symmetric and long-lived.  Each
filament is associated with a flux tube and can be well separated from the
surrounding turbulence.  Sheets exist in this regime as well, and they are
localized to the interface between flux tubes.  With $\eta$ small and $\mu =
0$, the system is in a sheet-dominated regime.  Both regimes have density
gradients that are non-Gaussian with large kurtosis.

The effects on pulsar signal scintillation in each regime have yet to be
ascertained directly.  The conventional picture of a L\'evy flight is a random
walk with step sizes distributed according to a long-tailed distribution with
no defined variance.  This gives rise to long, uninterrupted flights punctuated
by large scattering events. This is in contrast to a normally-distributed
random walk with relatively uniform step sizes and small scattering events.
The intermittent filaments that arise in the small $\eta$ and $\mu$ regime are
suggestive of structures that could scatter pulsar signals through large
angles, however the associated density structures are broadened in comparison
to the current filament and would not give rise to as large a scattering event.
Even broadened structures can yield L\'evy distributed density gradients
\citep{terry-smith07}, but it is not clear how the L\'evy flight picture can be
applied to these broad density gradient structures.  In the $\mu = 0$ regime,
the large-aspect-ratio sheets may serve to provide the necessary scatterings
through refraction and may map well onto the L\'evy flight model.

An alternative possibility, suggested by the \emph{temporal} intermittency of
the kurtosis (itself a measure of a field's \emph{spatial} intermittency), is
the encounter between the pulsar signal and a short-lived sheet that arises
during the merger of two filamentary structures.  These sheets are limited in
extent and have very large amplitudes.  At their greatest magnitude they are
the dominant structure in the numerical solution.  Their temporal intermittency
distinguish them from the long-lived sheets surrounding them.  It is possible
that a pulsar signal would undergo large scattering when interacting with a
merger sheet. This scattering would be a rare event, suggestive of a scenario
that would give rise to a L\'evy flight.
%}}}

\section{Acknowledgments} %{{{

We thank S. Boldyrev, S. Spangler and E. Zweibel for helpful discussions and
comments.

%}}}

%}}}

% Figures:%{{{

\clearpage

% \begin{figure}
% \includegraphics[width=\textwidth]{dengrad_x-0000000-uncorrelated-pdf.eps}
% \caption{log-PDF of density gradients (blue line) for an ensemble simulation
% with $\eta / \mu =
% 1$ at $t=0$.  The density gradient field is Gaussian distributed.  A best-fit
% Gaussian PDF is plotted for comparison in green.}
% \label{fig:dengrad_x-0000000-uncorrelated}
% \end{figure}

% \begin{figure}
% \includegraphics[width=\textwidth]{dengrad_x-0005000-uncorrelated-pdf.eps}
% \caption{log-PDF of density gradients (blue line) for an ensemble simulation
% with $\eta / \mu = 1$ at $t=5000$ simulation times.  The density gradient field is
% non-Gaussian.  A best-fit Gaussian PDF is plotted for comparison in
% green.  Non-Gaussian tails are evident.} 
% \label{fig:dengrad_x-0005000-uncorrelated} 
% \end{figure}

% \begin{figure}
% \includegraphics[width=\textwidth]{dengrad_x-0000000-zero-diff-pdf.eps}
% \caption{log-PDF of density gradients (blue line) for an ensemble simulation
% with $\eta / \mu \ll 1$ at $t=0$ simulation times.  The density gradient field is
% Gaussian distributed.  A best-fit Gaussian PDF is plotted for comparison in
% green.}  
% \label{fig:dengrad_x-0000000-zero-mu}
% \end{figure}

% \begin{figure}
% \includegraphics[width=\textwidth]{dengrad_x-0003000-zero-diff-pdf.eps}
% \caption{log-PDF of density gradients (blue line) for an ensemble simulation
% with $\eta / \mu \ll 1$ at $t=3000$ simulation times.  The density gradient field is
% non-Gaussian.  A best-fit Gaussian PDF is plotted for comparison in
% green.  Non-Gaussian tails are evident.}
% \label{fig:dengrad_x-0003000-zero-mu}
% \end{figure}

%}}}

\end{document}